\documentclass[
reprint,            
twocolumn,
superscriptaddress, 
amsmath,            
amssymb,            
aps,                
prd,                 
notitlepage,        
floatfix,           
nofootinbib,        
]{revtex4-1}

\usepackage{tensor}     
\usepackage{graphicx}   
\usepackage[
colorlinks=true,        
citecolor=blue,         
linkcolor=blue,         
urlcolor=blue           
]{hyperref}             
\usepackage{bm}         
\usepackage{xcolor}     
\usepackage{color}      
\usepackage[section]{placeins} 

\newcommand{\nc}{\newcommand*}

\def\bea{\begin{equation}}
\def\eea{\end{equation}}
\def\beq{\begin{eqnarray}}
\def\eeq{\end{eqnarray}}

\nc{\al}{\alpha}
\nc{\s}{\sigma}
\nc{\dt}{\delta}
\nc{\Dt}{\Delta}
\nc{\Ld}{\Lambda}
\nc{\p}{\partial}
\nc{\om}{\omega}
\nc{\Om}{\Omega}
\nc{\rd}{\mathrm{d}}
\nc{\Od}[1]{\mathcal{O}(#1)} 
\nc{\kp}{\kappa}

\def\({\left(}
\def\){\right)}
\def\[{\left[}
\def\]{\right]}
\def\e{\begin{equation}}
\def\q{\end{equation}}
\def\m{\begin{eqnarray}}
\def\n{\end{eqnarray}}

\nc{\Eq}[1]{Eq.~\eqref{#1}}     
\nc{\Fig}[1]{Fig.~\ref{#1}}     
\nc{\Table}[1]{Table~\ref{#1}}  
\nc{\Sec}[1]{Sec.~\ref{#1}}     

\nc{\Msun}{M_\odot}             
\nc{\fpbh}{f_{\mathrm{PBH}}}    
\nc{\fpbhn}{f_{\mathrm{PBH0}}}    
\nc{\mR}{\mathcal{R}} 
\nc{\seq}{\sigma_{\mathrm{eq}}}
\nc{\ogw}{\Omega_{\mathrm{GW}}}
\nc{\gpcyr}{\mathrm{Gpc}^{-3}\,\mathrm{yr}^{-1}}
\nc{\lvc}{LIGO/Virgo} 
\nc{\SNR}{\mathrm{SNR}} 
\nc{\mmin}{{m_{\mathrm{min}}}}
\nc{\mmax}{{m_{\mathrm{max}}}}
\nc{\Mmin}{{M_{\mathrm{min}}}}
\nc{\fmin}{{f_{\mathrm{min}}}}
\nc{\VT}{\mathrm{VT}}
\nc{\rhoGW}{\rho_{\mathrm{GW}}}
\nc{\vth}{\vec{\theta}}
\nc{\vd}{\vec{d}}
\nc{\vla}{\vec{\lambda}}
\nc{\Nobs}{N_{\mathrm{obs}}}
\nc{\av}[1]{\langle #1 \rangle} 
\nc{\km}{\mathrm{km}}
\nc{\Mpc}{\mathrm{Mpc}}
\nc{\Tobs}{T_{\mathrm{obs}}}
\nc{\Ntemp}{N_{\mathrm{temp}}}

\nc{\addref}{[\textcolor{red}{add ref}] } 
\nc{\eg}{\textit{e.g.~}}
\nc{\app}{\approx}
\nc{\hf}{\frac{1}{2}}
\nc{\discuss}{\textcolor{red}{Add discussion here!}}
\nc{\red}[1]{\textcolor{red}{#1}}

\nc{\mH}{\mathcal{H}}
\nc{\cs}{c_s^2}
\nc{\Sij}[1]{S_{ij}^{(#1)}}
\nc{\vi}[1]{v_i^{(#1)}}
\nc{\no}{\nonumber}
\def\<{\left\langle}
\def\>{\right\rangle}

\nc{\bk}{\bm{k}}
\nc{\bq}{\bm{q}}
\nc{\bp}{\bm{p}}
\nc{\bl}{\bm{l}}
\nc{\bx}{\bm{x}}
\nc{\be}{\mathbf{e}}
\nc{\mS}{\mathcal{S}}
\nc{\te}{\tilde{\eta}}
\nc{\tp}{\tilde{p}}
\nc{\tk}{\tilde{k}}
\nc{\tx}{\tilde{x}}
\nc{\tF}{\tilde{F}}
\nc{\tA}{\tilde{A}}
\nc{\mkpq}{|\bk-\bp-\bq|}
\nc{\mpq}{|\bp-\bq|}
\nc{\mkp}{|\bk-\bp|}
\nc{\mSi}[1]{\mS^{(#1)}({\bk, \eta})}
\nc{\mbh}{m_{\rm bh}}
\nc{\dEdf}{{dE\over df}}

\begin{document}
	
	\title{Evaporating black holes:\\ constraints on anomalous emission mechanisms}
	
	\author{Chen Yuan}
	\email{yuanchen@itp.ac.cn}
	\affiliation{CAS Key Laboratory of Theoretical Physics, 
		Institute of Theoretical Physics, Chinese Academy of Sciences, Beijing 100190, China}
	\affiliation{School of Physical Sciences, University of Chinese Academy of Sciences, No. 19A Yuquan Road, Beijing 100049, China}
	\author{Richard Brito}
	\affiliation{Dipartimento di Fisica, ``Sapienza" Università di Roma \& Sezione INFN Roma1, Piazzale Aldo Moro 5, 
		00185, Roma, Italy}
	\affiliation{CENTRA, Departamento de F\'{\i}sica, Instituto Superior T\'ecnico -- IST, Universidade de Lisboa -- UL,
		Avenida Rovisco Pais 1, 1049 Lisboa, Portugal}
	\author{Vitor Cardoso}
	\affiliation{CENTRA, Departamento de F\'{\i}sica, Instituto Superior T\'ecnico -- IST, Universidade de Lisboa -- UL,
		Avenida Rovisco Pais 1, 1049 Lisboa, Portugal}
	
	\date{\today}
	\begin{abstract}
		Hawking radiation of astrophysical black holes is minute and thought to be unobservable. 
		However, different mechanisms could contribute to an anomalously high emission rate:
		extra dimensions, new ``dark'' families of bosons or fermions, or a lower fundamental Planck scale.
		Do black holes flood the Universe with gravitational waves via mass loss? Here, we show that
		the formation of black hole binaries and the absence of a stochastic background of gravitational waves can limit the emission rate to
		$|\dot{M}|\lesssim 10^{-15}M_{\odot}/{\rm yr}$ ($|\dot{M}|\lesssim 10^{-13}M_{\odot}/{\rm yr}$), when the mass loss branching ratio to gravitons is unity ($10^{-2}$). This constraint is up to seven orders of magnitude more stringent than bounds from resolvable inspiralling binaries.
	\end{abstract}

	\maketitle

	\section{Introduction}
	Quantum field theory in curved spacetimes predicts that black holes (BHs) evaporate, with an equivalent temperature~\cite{Hawking:1974sw,Birrell:1982ix}
	\bea
	T=\frac{\hbar c^3}{8\pi G k_B M}\sim 6\times 10^{-8}\frac{M_\odot}{M}\,{\rm K}\,,
	\eea
	for a BH of mass $M$.	Four fundamental constants of nature are at work: Planck's constant $\hbar$, the speed of light $c$, Newton's constant $G$
	and Boltzmann's constant $k_B$. Their appearance in BH evaporation is an important hint of the foundational character of this process,
	which brings together Quantum Mechanics and General Relativity in the strong-field regime.
	The Hawking temperature above is too small to have any meaningful effect on BHs formed via gravitational collapse of stars.
	To see this, note that BHs evaporate with a power~\cite{Page:1976df}
	\bea
	\dot{M}=-\frac{k_B^4}{\hbar^3c^2}T^4\sigma\sim -10^{-30}\frac{M_\odot^2}{M^2}\,{\rm J/s}\,,
	\eea
	where $\sigma \propto G^2M^2/c^4$ is an effective cross section. Thus a stellar mass BH takes over $10^{60}$ years to evaporate~\cite{Page:1976df}.

	The predictions above can be challenged in different ways. It was noted, for example, that scenarios where the gravitational interaction
	is naturally higher-dimensional are consistent with observations~\cite{ArkaniHamed:1998rs,Antoniadis:1998ig,Randall:1999ee}.
	Compactified extra dimensions are -- on general grounds -- expected to have a compactification scale of the order of the Planck length $l_{\rm P}=\sqrt{\hbar G/c^3}\sim10^{-33}$
	cm. However, large or warped extra dimensions\,\cite{ArkaniHamed:1998rs,Antoniadis:1998ig,Randall:1999ee}, with a compactification scale $L\gg l_{\rm P}$, could (apparently) solve some of the puzzles of the standard model of particle physics, such as the
	longstanding hierarchy problem (i.e., the enormous gap between the electroweak energy scale and the Planck energy
	$E_{\rm P}\sim10^{19}$ GeV) could be addressed, since the effective Planck energy would be lowered.
	Laboratory tests of gravity verified the inverse-square law down to the micrometer scale~\cite{Sushkov:2011zz},
	but there is still plenty of room for new physics. Since larger energy scales are unreachable by present detectors - setting stronger
	constraints is a challenging programme. In higher dimensional spacetimes, the inverse square law is changed, leading to a different dynamics for compact binaries.
	In fact, for well-motivated setups, exciting the Kaluza-Klein modes is extremely difficult on astrophysical scales~\cite{Cardoso:2019vof}, but naive extra dimensional models can still be constrained~\cite{Pardo:2018ipy,Abbott:2018lct}. It is expected that gravity becomes ``stronger'' at shorter scales in these frameworks, possibly leading to copious Hawking radiation.
	However, to understand BH evaporation one needs to understand BHs themselves, their topology and possible phase space and
	semiclassical quantization of fields in such backgrounds. This program alone is also difficult, a challenge illustrated by the wealth of solutions in higher dimensional spacetimes with non-compact dimensions~\cite{Dias:2009iu,Emparan:2009at}.\footnote{Our ignorance on all possible BH solutions
		in higher dimensions, has led to the erroneous assumption that these BHs would evaporate extraordinarily fast, leading to signatures in the inspiral of two of these objects~\cite{McWilliams:2009ym}. The underlying hypothesis has since been disproved with the construction of new BH solutions in warped spacetimes~\cite{Figueras:2011gd,Abdolrahimi:2012qi}.}

	But higher-dimensions are not the only possible way to enhance BH evaporation. New families of light particles will increase the evaporation rate. These could be all or just a fraction of dark matter, and could include a wealth of new fields~\cite{Dine:1982ah,Dvali:2007hz,Arvanitaki:2009fg,Davoudiasl:2020uig}. Attempts at reconciling unitary evolution of Quantum Mechanics with BH evaporation make use of horizon-scale modifications to the geometry, which could also lead to an anomalous energy emission from BHs~\cite{Giddings:2017mym,Giddings:2017jts}.
	Finally, recent attempts to introduce gravitation into the framework
	of trace dynamics pre-quantum mechanics~\cite{Adler:2002fu}, assuming the metric to be described as usual by a classical field, led to new BH-like spacetimes~\cite{Adler:2013mna}.
	These spacetimes may give rise to large ``BH winds'' and consequent mass loss~\cite{Adler:2021urw}.

	\section{Constraints on black hole evaporation}
	\subsection{Binary-inspiral limits}
	What exactly {\it are} the constraints on BH evaporation? For isotropic emission in the rest frame, with
	\bea
	\dot{M}=-\alpha\,,\label{mass_loss_model1}
	\eea
	the impact of mass-loss on the evolution of a binary was studied both in the context of stars and BHs~\cite{1963Icar....2..440H,1966ZA.....63..116H,Simonetti:2010mk,McWilliams:2009ym,Chung:2020uqj}. The effects of mass loss enter at $-4$ PN order in the expansion of the gravitational-wave (GW) phase~\cite{Yagi:2011yu,Perkins:2020tra}. Thus, precision GW astronomy can help in constraining such mechanism. The most stringent forecasted bounds can be obtained with BH binaries that would be detectable by both future, low-frequency space-based GW observatories and ground-based detectors, and yield~\cite{Perkins:2020tra}
	\bea
	\alpha_{\rm inspiral}\lesssim 10^{-8}\frac{M_{\odot}}{\rm yr}\,.\label{inspiral}
	\eea

		\subsection{Bounds from existence of BHs and BH binaries}
	The above best possible bounds from the observation of compact binaries in the GW window
	can be improved in different ways. The mere existence of stellar-mass BHs, now firmly established, indicates that BHs can survive across a Hubble time. For instance, constraints based on the age of stellar-mass BHs were obtained for the BH in the binary system XTE J1118+480, with an estimated age $\gtrsim11$Myr \cite{Psaltis:2006de}, and for the BH candidate in the globular cluster RZ2109, which has an estimated mass $\sim 10\Msun$ and age $\sim 10^{10}\mathrm{yr}$ \cite{Gnedin:2009yt,steele2014composition}.
	Thus,
	\bea\label{existence}
	\alpha_{\rm existence BH}\lesssim 10^{-10}\frac{M_{\odot}}{\rm yr}\,,
	\eea
	already better than the previous bound \eqref{inspiral}, inspired on binary evolution.

	Mass loss in a binary leads generically to an outspiral~\cite{1963Icar....2..440H,1966ZA.....63..116H,Simonetti:2010mk,McWilliams:2009ym}.
	Thus, a constant mass loss that dominates over GW emission will prevent BHs from merging. LIGO/Virgo observations therefore impose limits on mass loss from BHs.
	For the isotropic mass loss~\eqref{mass_loss_model1}, the specific angular momentum is conserved, and one finds the evolution law for the semi-major axis 
	\bea
	\dot{a}\sim -a\frac{\dot{M}}{M}\,,
	\eea
	where $M$ is the binary's total mass. For simplicity we focus on equal-mass binaries and we use units where the speed of light $c$ and Newton's constant $G$ are $G=c=1$.
	On the other hand, GW emission leads to~\cite{Peters:1964zz}
	\bea
	\dot{a}\sim -\frac{16}{5}\frac{M^3}{a^3}\,.
	\eea
	Thus, balance between outspiral and inspiral can be estimated to occur at a mass loss
	\bea
	\dot{M}\approx \left(\frac{M}{a}\right)^4=\frac{c^3}{G}\left(\frac{GM}{c^2a}\right)^4\,,
	\eea
	leading to the constraint
	\bea\label{merge_const}
	\alpha_{\rm BH coalescence}\lesssim 10^{-15}\,\left(\frac{10^7 M}{a}\right)^4\,{M_{\odot}\over \mathrm{yr}}\,.
	\eea
	We normalized the distance by what could be characteristic values for a newly-born compact binary~\cite{Belczynski:2001uc,Barack:2018yly}.
	There is a great deal of uncertainty here, since common envelope physics leads to dynamical friction which could overtake mass loss. 
	Nevertheless, most of the coalescences of stellar-mass BHs are consistent with zero or very small eccentricities which limit such mechanisms~\cite{Cardoso:2020iji}. On the other hand, accretion effects can contribute to a mass increase which can counterbalance the mass loss mechanism. If one considers a binary BH system formed in isolation, a conservative upper limit for Eq.~(\ref{merge_const}) comes from considering the typical accretion rate of interstellar medium for isolated solar mass BHs, given by $\sim10^{11}\Msun/\mathrm{yr}$~\cite{Agol:2001hb}, therefore yielding $\alpha_{\rm BH coalescence}\lesssim 10^{-11}\Msun/\mathrm{yr}$ as a conservative estimate.

\subsection{Bounds from stochastic radiation}
An equally stringent bound can be derived from the absence of stochastic gravitational radiation. Unlike in the constraints discussed above, where the BH mass loss could be due to the emission of any kind of particles, here we will {\it assume} that there is some mechanism (for example, one of those discussed in the introduction) whereby BHs can have an anomalously high energy loss, likely of quantum nature, dominated by the emission of gravitons. Given the only scale of the problem -- the BH mass $M$ -- we take the GW emission to be dominated by gravitons which in the source frame have a frequency
\bea\label{fm}
f_s \approx 32.3 \mathrm{kHz}\({M\over \Msun}\)^{-1}\,.
\eea

If BHs do have an intrinsic mass loss rate $\dot{M}$ due to the emission of gravitons, then the incoherent superposition of unresolvable GWs from a BH population can produce a stochastic GW background (SGWB). See, e.g., \cite{Dong:2015yjs} where the GW background due to standard Hawking radiation from a population of primordial BHs  was computed.  Let us therefore estimate this SGWB and explore its detectability with LIGO~\cite{TheLIGOScientific:2014jea} and future GW detectors such as the Neutron Star Extreme Matter Observatory (NEMO)~\cite{Ackley:2020atn}, Cosmic Explorer (CE)~\cite{Evans:2016mbw} and the Einstein Telescope (ET) \cite{Punturo:2010zz,Maggiore:2019uih}.

Let $dE_s/df_s$ be the GW energy spectrum in the source frame. Now, consider the energy emitted in the frequency band $[f_0,f_0+\Delta f]$,
\e\label{energy}
\int_{f_0}^{f_0+\Delta f}{dE_s\over df_s'}df_s'\approx {dE_s\over df_s}\Delta f\,,
\q
where we assumed $\Delta f/f_0\ll 1$. Although multi-metric theories (e.g., \cite{Hassan:2012wt,Hinterbichler:2012cn}) could in principle predict the enhancement of BH evaporation, their coupling to matter can be complicated (e.g., \cite{Noller:2014sta}). As a result, it is quite unclear the fraction of radiation that goes into GWs. Let $P_g$ be the fraction of energy a BH loses through the emission of gravitons. For standard Hawking radiation $P_g\sim 0.01$~\cite{Page:1976df}. Then we can relate Eq.~(\ref{energy}) to the mass loss rate $\dot{M}$ using
\e\label{EfEdot}
{dE_s\over df_s} \Delta f =P_g |\dot{M}| \Delta t \,,
\q
where $\Delta t = t(M_2) - t(M_1)$ is the time that the BH evolves from having a mass $M_1$ to a mass $M_2$, and $t(M)$ can be obtained by solving $\dot{M}$, as we do below for specific models. 
Combining the above equations and taking the limit $\Delta f\to 0$ (which also implies $\Delta t\to 0$), the GW energy spectrum can then be expressed as
\e\label{gdEdf}
{dE_s\over df_s}=P_g|\dot{M}|{d t \over df_s}, ~ f_s\in[f_{\min},f_{\max}]\,,
\q
where $f_{\min}$ is the frequency of the GWs emitted by the BH when it was born with mass $M_i$, whereas $f_{\max}$ is the GW frequency emitted in the present time. For example, if the BH has evaporated by today, then $f_{\max}\to +\infty$, otherwise $f_{\max}$ can be computed by evolving the BH mass from its formation time to the present. Note also that, within our assumptions, there is a one-to-one correspondence between the BH mass, the time since BH formation and the source frame GW frequency, i.e. $M\equiv M(f_s), t\equiv t(f_s)$.

To compute the energy spectrum~\eqref{gdEdf} we now consider two different models for the mass loss. The first model is the same agnostic model considered above: a constant mass loss rate given by Eq.~\eqref{mass_loss_model1}, which predicts that the BH mass evolves as
\e\label{t1}
M(t)=M_i-\alpha t \,.
\q
The second model takes the same form of mass loss rate as the one caused by Hawking radiation~\cite{Page:1976df},
\e
\dot{M}=-{\alpha_H\over M^2}\,,
\q
whose solution is given by
\e\label{t2}
M(t)=\(M_i^3-3\alpha_H t\)^{1/3}\,.
\q
For both models, using Eq.~\eqref{gdEdf}, we find that the GW energy spectrum reads
\e\label{pdEdf}
{dE_s\over df_s}={P_g\over 2\pi f_s^2}\,, ~f_s\in[f_{\min},f_{\max}]\,.
\q
In fact, one can easily check that for a power law model, $\dot{M}\propto-M^n$ with $n$ an arbitrary real number, the GW energy spectrum is always given by Eq.~\eqref{pdEdf}.

To characterize the SGWB we compute the dimensionless energy density parameter, $\Omega_{\mathrm{GW}}(f)$, defined as
\e
\Omega_{\mathrm{GW}}(f)\equiv {1\over \rho_c}\frac{d\rho_{\mathrm{GW}}}{d \ln(f)}\,,
\q
where $\rho_c=3H_0^2/(8\pi)$ is the critical energy of the Universe.
Assuming that the GW sources are isotropically distributed in the sky, the density parameter can be computed by summing over all the sources~\cite{Phinney:2001di},
\e\label{Ogw}
\Omega_{\mathrm{gw}}(f)=\frac{f}{\rho_{c}} \int d M d z \frac{d t}{d z} \frac{d \dot{n}}{d M} \frac{d E_{s}}{d f_{s}}\,,
\q
where $dt/dz$ stands for the derivative of the lookback time with respect to the cosmological redshift. The frequency in the detector frame is related to the source frame frequency by $f_s=f(1+z)$. In this paper, we focus on extragalactic BHs formed from stellar collapse whose formation rate per co-moving volume is given by \cite{Brito:2017wnc,Tsukada:2018mbp,Tsukada:2020lgt,Yuan:2021ebu}:
\e\label{ndot}
\frac{d \dot{n}(z)}{d M}= \int \psi[t-\tau(M_*)]\phi(M_*)\delta(M_*-g^{-1}_{\mathrm{rem}})\mathrm{d}M_*\,,
\q
where $\delta$ is the Dirac delta function. Here, $\tau(M_*)$ describes the lifetime of the progenitor star for a given mass $M_*$~\cite{Schaerer:2001jc}. $\phi(M_*)\propto M_*^{-2.35}$ is the Salpeter initial mass function of the progenitor stars, and we normalized it in the range $[0.1,100]\Msun$. The function $g_{\mathrm{rem}}=g_{\mathrm{rem}}(M_*,z)$ returns the mass of the BH remnant for a given $M_*$ at redshift $z$~\cite{Fryer:2011cx} and $g^{-1}_{\mathrm{rem}}$ is the inverse function of $g_{\mathrm{rem}}$. The $z$ dependence of $g_{\mathrm{rem}}(M_*,z)$ comes from the metallicity of the BH progenitor for which we adopt the results in~\cite{Belczynski:2016obo} and we take $Z_\odot=0.0196$~\cite{Vagnozzi:2017wge}. The metallicity as a function of $z$ that we consider is only valid if $z\in[0,20]$ and thus we will only consider BHs formed within this range. For the cosmic star formation rate $\psi(t)$, we adopt the functional form in \cite{Springel:2002ux} and the fitted parameters given in \cite{Vangioni:2014axa}. Finally, we set the BH mass to be in the range $M\in[3,50]\Msun$. 

\begin{figure}[htbp!]
	\centering
	\includegraphics[width = 0.5\textwidth]{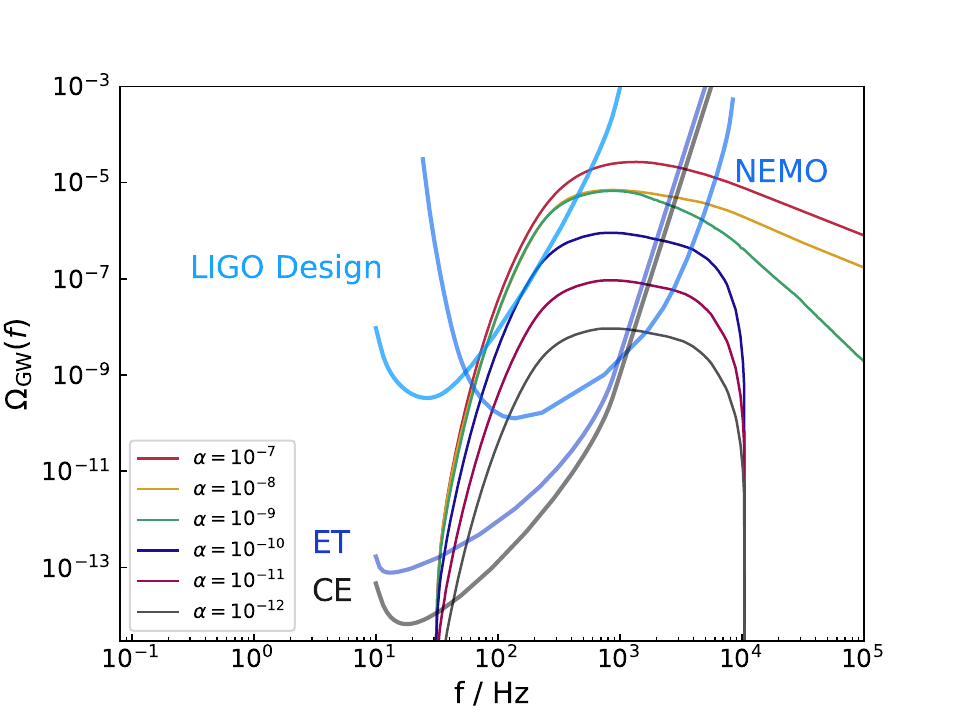}
	\includegraphics[width = 0.5\textwidth]{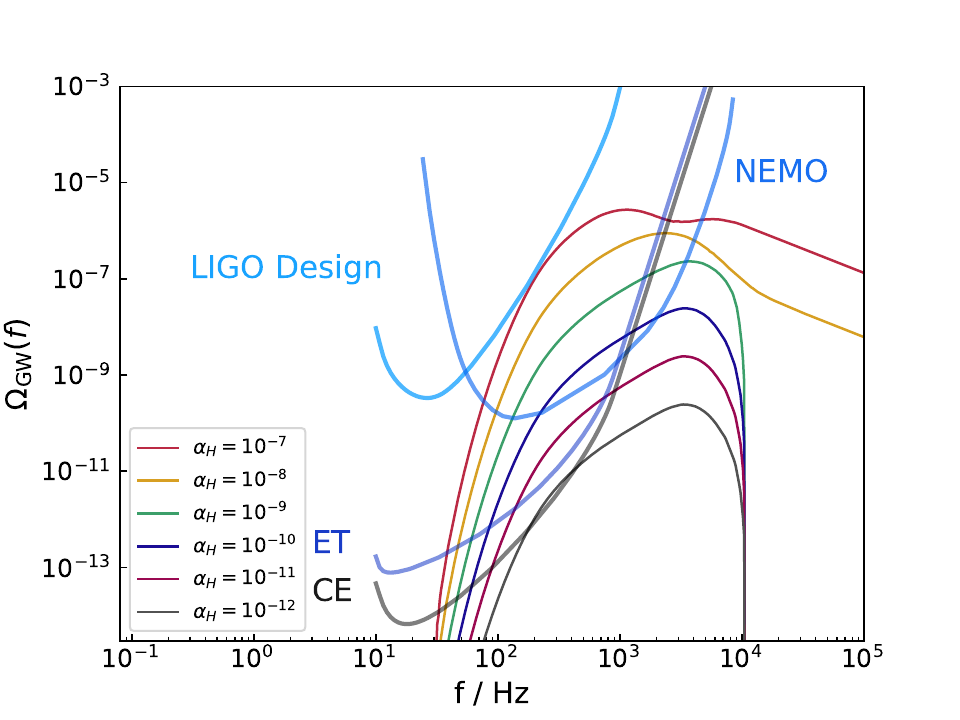}
	\caption{The SGWB from isolated stellar-origin BHs due to quantum effects with $P_g=1$.. Upper panel:  $\alpha$ is in units $\Msun/\mathrm{yr}$. Lower Panel:  $\alpha_H$ is in units $\Msun^3/\mathrm{yr}$. The power-law integrated sensitivity curves~\cite{Thrane:2013oya} for LIGO/NEMO/CE/ET are shown at their design sensitivity, all corresponding to $\SNR=1$ and a four-year-observation time. We assume two co-aligned and co-located identical detectors for NEMO/CE/ET and we compute the overlap function for LIGO based on~\cite{Thrane:2013oya}.}\label{ogw}
\end{figure}
The SGWB obtained with both models, along with the sensitivity curves of different GW detectors, are shown in Fig.~\ref{ogw}.  For concreteness, we set $P_g=1$. Since $\Omega_\mathrm{GW}\propto P_g$, the reader can easily estimate $\Omega_{\mathrm{GW}}$ for different values of $P_g$ by just multiplying the results shown in Fig.~\ref{ogw} by the desired value of $P_g$.
Some features of the GW background are quite generic, and can be easily understood. The lower frequency cutoff of $\Omega_{\mathrm{GW}}$ corresponds to GWs emitted by the farthest and largest BHs in the population, i.e. BHs born with mass $M_i\sim 50\Msun$ at redshift $z=20$, that emit mainly at frequencies $\sim 32.3\mathrm{kHz}\,\Msun/\left((1+z)M_i\right) \simeq 30$Hz. 
The upper frequency cutoff instead depends on the magnitude of the BH mass loss rate. From Eqs.~\eqref{t1}--\eqref{t2}, we can see that for a BH to completely evaporate by today one needs $\alpha>M_i/t_a$ ($\alpha_H>M_i^3/(3t_a)$) where $t_a\sim 10^9$ yr is the typical age of the BHs in the population. Therefore for sufficiently large $\alpha$, $f_{\max}$ can be infinitely large (e.g., the red and the yellow lines in Fig.~\ref{ogw}). On the other hand, in the small $\alpha$ limit, $M$ is nearly constant and thus the upper frequency cutoff is set by the less massive BHs in the population, $M_i=3\Msun$, born at small cosmological redshifts $z\sim 0$. This gives an upper frequency $32.3\mathrm{kHz}\,\Msun/\left((1+z)M_i\right) \simeq 10.8$kHz, in agreement with what we find in Fig.~\ref{ogw}.
Furthermore, we note that in the small $\alpha$ limit, the BH mass can be regarded as a constant and in this case, $dE_s/df_s$ approaches a $\delta$-function with amplitude $\simeq \dot{M}t_a\propto \alpha$. Thus, we have $\Omega_{\mathrm{GW}} \propto \alpha$, also in agreement with our results.
Finally, we note that small changes of the source-frame frequency, Eq.~\eqref{fm}, would only slightly shift the lower and higher cutoff of $\Omega_{\mathrm{GW}}$ while its amplitude would be almost unchanged.
%

To quantitatively estimate the prospects to constrain such a background, we calculate the signal-to-noise ratio (SNR) for the different GW detectors for which we show power-law sensitivity curves in Fig.~\ref{ogw}. The SNR for an arbitrary large SGWB, can be computed using~\cite{Allen:1997ad,Yuan:2021ebu}:
\bea
\rho^{2}=T  \int  \mathrm{d} f \frac{\Gamma^{2} S_{h}^{2}}{\left[{1\over 25}+\Gamma^{2}\right] S_{h}^{2}+P_{n}^{2}+{2\over 5} S_{h} P_{n}}\,,
\eea
where $\Gamma(f)$ is the overlap function and $P_n(f)$ is the noise power spectral density of the detector. The strain power spectral density, $S_h(f)$, is related to $\Omega_{\mathrm{GW}}(f)$ by $S_h(f) = 3H_0^2\Omega_{\mathrm{GW}}(f)/(2\pi^2 f^3)$. We take the observation time to be $T=4$ yr and take the same detector configurations considered in~\cite{Yuan:2021ebu} and also described in Fig.~\ref{ogw}.

\begin{figure}[htbp!]
	\centering
	\includegraphics[width = 0.5\textwidth]{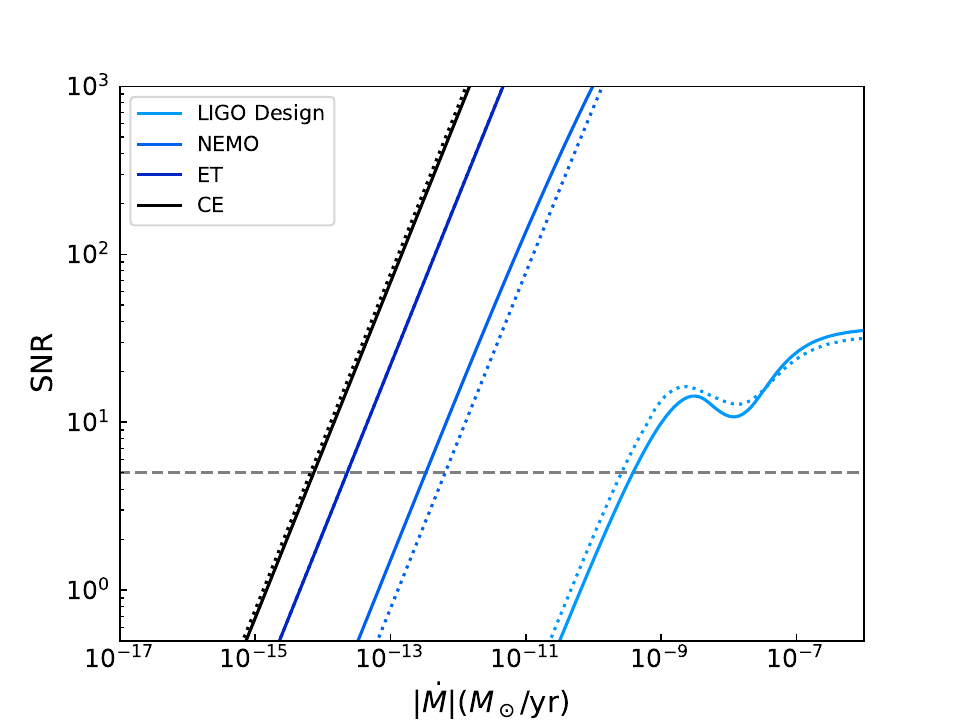}
	\includegraphics[width = 0.5\textwidth]{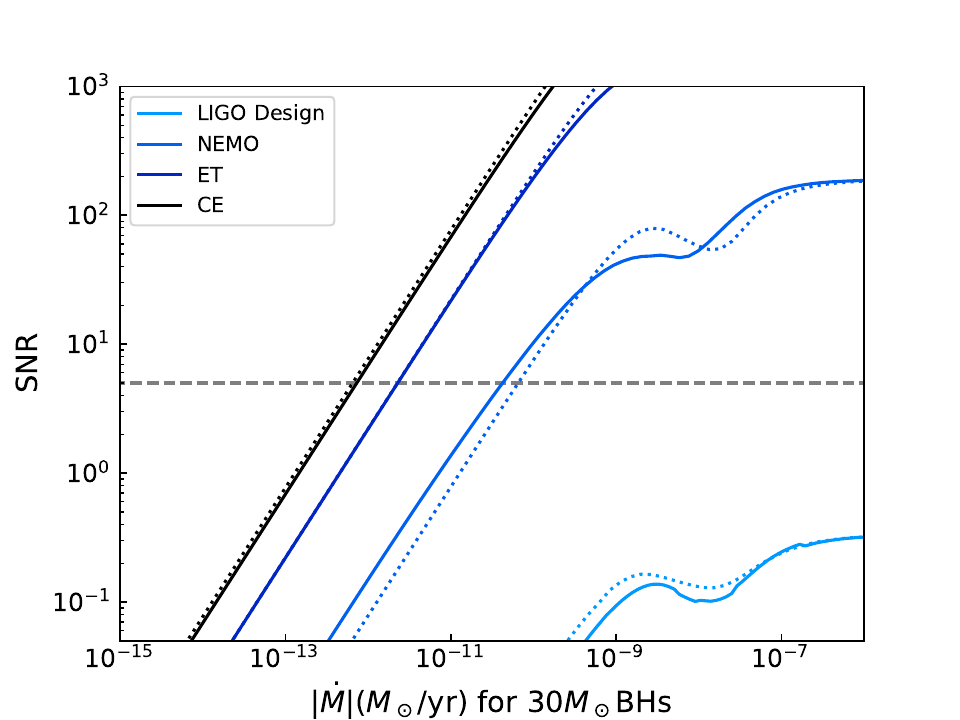}
	\caption{The expected SNR for the SGWB as a function of $|\dot{M}|$
		Upper panel: $P_g=1$. Lower panel: $P_g=0.01$.
		Solid lines: The expected SNR for $\dot{M}=-\alpha$.
		Dashed lines: The expected SNR for $\dot{M}=-\alpha_H/M^2$ where we fixed the mass of BHs to be $M=30\Msun$.}\label{SNR}
\end{figure}

As argued above, in the small $\alpha$ and $\alpha_H$ limit, the BH mass can be considered to be nearly constant over an Hubble time and the two mass loss models that we considered above are related by $\alpha=\alpha_H/M^2$. In this limit, the overall background amplitude is nearly model-independent, and is mostly set by the magnitude of $\dot{M}$. For concreteness and given that $\dot{M}$ can be a function of the BH mass, we focus on $|\dot{M}|$ for a typical value $M_i=30\Msun$.
The SNR for the SGWB is shown in Fig.~\ref{SNR} where the dashed line corresponds to $\mathrm{SNR}=5$ above which a detection of the SGWB could be claimed. The upper limits, below which $\mathrm{SNR}<5$, obtained with different detectors on $|\dot{M}|$ are listed in Table~\ref{tablealpha}.

\begin{table}[htbp!]
	\begin{center}
		\setlength{\tabcolsep}{0.7mm}{
			\begin{tabular}{c|cccc}
				\hline\hline
				model & LIGO & NEMO &  ET  & CE \\
				\hline
				$\mathcal{M},P_g=1$  & $2.6\!\times\!10^{-10}$ & $6.4\!\times\!10^{-13}$ & $2.3\!\times\!10^{-14}$ & $6.5\!\times\!10^{-15}$\\
				\hline
				$\mathcal{M}_H,P_g=1$ & 
				$3.8\!\times\!10^{-10}$ & $3.3\!\times\!10^{-13}$ & $2.3\!\times\!10^{-14}$ & $7.3\!\times\!10^{-15}$\\
				\hline
				$\mathcal{M},P_g=0.01$   & - & $6.5\!\times\!10^{-11}$ & $2.3\!\times\!10^{-12}$ & $6.5\!\times\!10^{-13}$\\
				\hline
				$\mathcal{M}_H,P_g=0.01$ & 
				- & $4.2\!\times\!10^{-11}$ & $2.3\!\times\!10^{-12}$ & $7.3\!\times\!10^{-13}$\\
				\hline
		\end{tabular}}
	\end{center}  
	\caption{The expected upper limit on $|\dot{M}|$ for $30\Msun$ BHs in the two different models considered in this paper. $\mathcal{M}$ refers to $|\dot{M}|=-\alpha$ and $\mathcal{M}_H$ refers to $|\dot{M}|=-\alpha_H/M^2$. Here we defined the upper limits as the values for which the background ${\rm SNR}=5$.
	}\label{tablealpha}
\end{table}
As shown in Table~\ref{tablealpha}, the upper limit on $|\dot{M}|$ is nearly independent on the particular model, which agrees with our previous argument. This also indicates that our results should be robust for other possible models for the mass loss rate $\dot{M}$.
Moreover, for the most optimistic case where $P_g=1$, it can be seen that even LIGO will already be able to place a very stringent constraint, about two orders magnitude stronger than the one that is forecasted to be obtained from the observation of GWs from BH binaries, Eq.~\eqref{inspiral}, and comparable to the constraints imposed from the observation of stellar-mass BHs, Eq.~\eqref{existence}. Future GW detectors, such as NEMO/ET/CE, will be able to place even a more stringent constraint on $|\dot{M}|$, namely $|\dot{M}|\lesssim 10^{-15}M_{\odot}/{\rm yr}$ when assuming $P_g=1$ (or $|\dot{M}|\lesssim 10^{-13}M_{\odot}/{\rm yr}$ for the less optimistic case where $P_g=0.01$), comparable to the most optimistic constraints that can be imposed from the observation of merging binary BHs, Eq.~\eqref{merge_const}. We remind, however, that the constraints due to the SGWB refer to mass loss in the gravitational channel only, unlike the other constraints discussed in this paper, which are independent of the channels through which the BH evaporates.

\section{Discussion}
There seems to be wide space for new phenomena in the context of BH physics and quantum mechanics~\cite{Bekenstein:1997bt,Bekenstein:1995ju,Giddings:2017mym,Giddings:2017jts,Agullo:2020hxe}, some of them predicting much larger emission rates from BHs, as compared to Hawking's original calculation~\cite{Hawking:1974sw}. Here we argued that observations of BHs and GWs already limits the amount of energy that BHs in the stellar-mass range can lose through quantum processes to be smaller than $|\dot{M}| \sim 10^{-15} M_{\odot}/{\rm yr}$ for the most optimistic case (or $|\dot{M}|\lesssim 10^{-13}M_{\odot}/{\rm yr}$ for the less optimistic case where $P_g=0.01$).
This limit, inferred from the observation of merging BH binaries and from the absence of a stochastic background of GWs, is still many orders of magnitude above the expected Hawking mass loss.
In other words, the capability of current, or planned, detectors is orders of magnitude away from that necessary to probe standard Hawking radiation.
Nevertheless, it is several orders of magnitude more stringent than previous forecasted bounds. It is quite remarkable that several decades after Hawking first predicted BH evaporation,
with GW astronomy we now have strong observational evidence that, if astrophysical BHs do evaporate, they must do it in a very slow fashion. Not surprisingly, the constraints we obtained are still entirely consistent with Hawking's prediction, however they severely limit anomalously high evaporation rates. 

\section*{Acknowledgments}
We acknowledge the use of \texttt{GWSC.jl} package \cite{GWSC} in calculating the sensitivity curves.
R.B. acknowledges financial support provided under the European Union's H2020 ERC, Starting Grant agreement no.~DarkGRA--757480 and under the MIUR PRIN and FARE programmes (GW-NEXT, CUP:~B84I20000100001). R.B. also acknowledges financial support provided by FCT under the Scientific Employment Stimulus - Individual Call - 2020.00470.CEECIND.
V.C. acknowledges financial support provided under the European Union's H2020 ERC 
Consolidator Grant ``Matter and strong-field gravity: New frontiers in Einstein's 
theory'' grant agreement no. MaGRaTh--646597.
This project has received funding from the European Union's Horizon 2020 research and innovation programme under the Marie Sklodowska-Curie grant agreement No 101007855.
We thank FCT for financial support through Project~No.~UIDB/00099/2020.
We acknowledge financial support provided by FCT/Portugal through grants PTDC/MAT-APL/30043/2017 and PTDC/FIS-AST/7002/2020.
The authors would like to acknowledge networking support by the GWverse COST Action 
CA16104, ``Black holes, gravitational waves and fundamental physics.''
We also acknowledge support from the Amaldi Research Center funded by the MIUR program ``Dipartimento di Eccellenza'' (CUP:~B81I18001170001).
	
\bibliography{./ref}

\begin{thebibliography}{64}%
\makeatletter
\providecommand \@ifxundefined [1]{%
 \@ifx{#1\undefined}
}%
\providecommand \@ifnum [1]{%
 \ifnum #1\expandafter \@firstoftwo
 \else \expandafter \@secondoftwo
 \fi
}%
\providecommand \@ifx [1]{%
 \ifx #1\expandafter \@firstoftwo
 \else \expandafter \@secondoftwo
 \fi
}%
\providecommand \natexlab [1]{#1}%
\providecommand \enquote  [1]{``#1''}%
\providecommand \bibnamefont  [1]{#1}%
\providecommand \bibfnamefont [1]{#1}%
\providecommand \citenamefont [1]{#1}%
\providecommand \href@noop [0]{\@secondoftwo}%
\providecommand \href [0]{\begingroup \@sanitize@url \@href}%
\providecommand \@href[1]{\@@startlink{#1}\@@href}%
\providecommand \@@href[1]{\endgroup#1\@@endlink}%
\providecommand \@sanitize@url [0]{\catcode `\\12\catcode `\$12\catcode
  `\&12\catcode `\#12\catcode `\^12\catcode `\_12\catcode `\%12\relax}%
\providecommand \@@startlink[1]{}%
\providecommand \@@endlink[0]{}%
\providecommand \url  [0]{\begingroup\@sanitize@url \@url }%
\providecommand \@url [1]{\endgroup\@href {#1}{\urlprefix }}%
\providecommand \urlprefix  [0]{URL }%
\providecommand \Eprint [0]{\href }%
\providecommand \doibase [0]{http://dx.doi.org/}%
\providecommand \selectlanguage [0]{\@gobble}%
\providecommand \bibinfo  [0]{\@secondoftwo}%
\providecommand \bibfield  [0]{\@secondoftwo}%
\providecommand \translation [1]{[#1]}%
\providecommand \BibitemOpen [0]{}%
\providecommand \bibitemStop [0]{}%
\providecommand \bibitemNoStop [0]{.\EOS\space}%
\providecommand \EOS [0]{\spacefactor3000\relax}%
\providecommand \BibitemShut  [1]{\csname bibitem#1\endcsname}%
\let\auto@bib@innerbib\@empty
\bibitem [{\citenamefont {Hawking}(1975)}]{Hawking:1974sw}%
  \BibitemOpen
  \bibfield  {author} {\bibinfo {author} {\bibfnamefont {S.~W.}\ \bibnamefont
  {Hawking}},\ }\href {\doibase 10.1007/BF02345020} {\bibfield  {journal}
  {\bibinfo  {journal} {Commun. Math. Phys.}\ }\textbf {\bibinfo {volume}
  {43}},\ \bibinfo {pages} {199} (\bibinfo {year} {1975})},\ \bibinfo {note}
  {[Erratum: Commun.Math.Phys. 46, 206 (1976)]}\BibitemShut {NoStop}%
\bibitem [{\citenamefont {Birrell}\ and\ \citenamefont
  {Davies}(1984)}]{Birrell:1982ix}%
  \BibitemOpen
  \bibfield  {author} {\bibinfo {author} {\bibfnamefont {N.~D.}\ \bibnamefont
  {Birrell}}\ and\ \bibinfo {author} {\bibfnamefont {P.~C.~W.}\ \bibnamefont
  {Davies}},\ }\href {\doibase 10.1017/CBO9780511622632} {\emph {\bibinfo
  {title} {{Quantum Fields in Curved Space}}}},\ Cambridge Monographs on
  Mathematical Physics\ (\bibinfo  {publisher} {Cambridge Univ. Press},\
  \bibinfo {address} {Cambridge, UK},\ \bibinfo {year} {1984})\BibitemShut
  {NoStop}%
\bibitem [{\citenamefont {Page}(1976)}]{Page:1976df}%
  \BibitemOpen
  \bibfield  {author} {\bibinfo {author} {\bibfnamefont {D.~N.}\ \bibnamefont
  {Page}},\ }\href {\doibase 10.1103/PhysRevD.13.198} {\bibfield  {journal}
  {\bibinfo  {journal} {Phys. Rev. D}\ }\textbf {\bibinfo {volume} {13}},\
  \bibinfo {pages} {198} (\bibinfo {year} {1976})}\BibitemShut {NoStop}%
\bibitem [{\citenamefont {Arkani-Hamed}\ \emph {et~al.}(1998)\citenamefont
  {Arkani-Hamed}, \citenamefont {Dimopoulos},\ and\ \citenamefont
  {Dvali}}]{ArkaniHamed:1998rs}%
  \BibitemOpen
  \bibfield  {author} {\bibinfo {author} {\bibfnamefont {N.}~\bibnamefont
  {Arkani-Hamed}}, \bibinfo {author} {\bibfnamefont {S.}~\bibnamefont
  {Dimopoulos}}, \ and\ \bibinfo {author} {\bibfnamefont {G.~R.}\ \bibnamefont
  {Dvali}},\ }\href {\doibase 10.1016/S0370-2693(98)00466-3} {\bibfield
  {journal} {\bibinfo  {journal} {Phys. Lett.}\ }\textbf {\bibinfo {volume}
  {B429}},\ \bibinfo {pages} {263} (\bibinfo {year} {1998})},\ \Eprint
  {http://arxiv.org/abs/hep-ph/9803315} {arXiv:hep-ph/9803315 [hep-ph]}
  \BibitemShut {NoStop}%
\bibitem [{\citenamefont {Antoniadis}\ \emph {et~al.}(1998)\citenamefont
  {Antoniadis}, \citenamefont {Arkani-Hamed}, \citenamefont {Dimopoulos},\ and\
  \citenamefont {Dvali}}]{Antoniadis:1998ig}%
  \BibitemOpen
  \bibfield  {author} {\bibinfo {author} {\bibfnamefont {I.}~\bibnamefont
  {Antoniadis}}, \bibinfo {author} {\bibfnamefont {N.}~\bibnamefont
  {Arkani-Hamed}}, \bibinfo {author} {\bibfnamefont {S.}~\bibnamefont
  {Dimopoulos}}, \ and\ \bibinfo {author} {\bibfnamefont {G.~R.}\ \bibnamefont
  {Dvali}},\ }\href {\doibase 10.1016/S0370-2693(98)00860-0} {\bibfield
  {journal} {\bibinfo  {journal} {Phys. Lett.}\ }\textbf {\bibinfo {volume}
  {B436}},\ \bibinfo {pages} {257} (\bibinfo {year} {1998})},\ \Eprint
  {http://arxiv.org/abs/hep-ph/9804398} {arXiv:hep-ph/9804398 [hep-ph]}
  \BibitemShut {NoStop}%
\bibitem [{\citenamefont {Randall}\ and\ \citenamefont
  {Sundrum}(1999)}]{Randall:1999ee}%
  \BibitemOpen
  \bibfield  {author} {\bibinfo {author} {\bibfnamefont {L.}~\bibnamefont
  {Randall}}\ and\ \bibinfo {author} {\bibfnamefont {R.}~\bibnamefont
  {Sundrum}},\ }\href {\doibase 10.1103/PhysRevLett.83.3370} {\bibfield
  {journal} {\bibinfo  {journal} {Phys. Rev. Lett.}\ }\textbf {\bibinfo
  {volume} {83}},\ \bibinfo {pages} {3370} (\bibinfo {year} {1999})},\ \Eprint
  {http://arxiv.org/abs/hep-ph/9905221} {arXiv:hep-ph/9905221 [hep-ph]}
  \BibitemShut {NoStop}%
\bibitem [{\citenamefont {Sushkov}\ \emph {et~al.}(2011)\citenamefont
  {Sushkov}, \citenamefont {Kim}, \citenamefont {Dalvit},\ and\ \citenamefont
  {Lamoreaux}}]{Sushkov:2011zz}%
  \BibitemOpen
  \bibfield  {author} {\bibinfo {author} {\bibfnamefont {A.~O.}\ \bibnamefont
  {Sushkov}}, \bibinfo {author} {\bibfnamefont {W.~J.}\ \bibnamefont {Kim}},
  \bibinfo {author} {\bibfnamefont {D.~A.~R.}\ \bibnamefont {Dalvit}}, \ and\
  \bibinfo {author} {\bibfnamefont {S.~K.}\ \bibnamefont {Lamoreaux}},\ }\href
  {\doibase 10.1103/PhysRevLett.107.171101} {\bibfield  {journal} {\bibinfo
  {journal} {Phys. Rev. Lett.}\ }\textbf {\bibinfo {volume} {107}},\ \bibinfo
  {pages} {171101} (\bibinfo {year} {2011})},\ \Eprint
  {http://arxiv.org/abs/1108.2547} {arXiv:1108.2547 [quant-ph]} \BibitemShut
  {NoStop}%
\bibitem [{\citenamefont {Cardoso}\ \emph {et~al.}(2019)\citenamefont
  {Cardoso}, \citenamefont {Gualtieri},\ and\ \citenamefont
  {Moore}}]{Cardoso:2019vof}%
  \BibitemOpen
  \bibfield  {author} {\bibinfo {author} {\bibfnamefont {V.}~\bibnamefont
  {Cardoso}}, \bibinfo {author} {\bibfnamefont {L.}~\bibnamefont {Gualtieri}},
  \ and\ \bibinfo {author} {\bibfnamefont {C.~J.}\ \bibnamefont {Moore}},\
  }\href {\doibase 10.1103/PhysRevD.100.124037} {\bibfield  {journal} {\bibinfo
   {journal} {Phys. Rev. D}\ }\textbf {\bibinfo {volume} {100}},\ \bibinfo
  {pages} {124037} (\bibinfo {year} {2019})},\ \Eprint
  {http://arxiv.org/abs/1910.09557} {arXiv:1910.09557 [gr-qc]} \BibitemShut
  {NoStop}%
\bibitem [{\citenamefont {Pardo}\ \emph {et~al.}(2018)\citenamefont {Pardo},
  \citenamefont {Fishbach}, \citenamefont {Holz},\ and\ \citenamefont
  {Spergel}}]{Pardo:2018ipy}%
  \BibitemOpen
  \bibfield  {author} {\bibinfo {author} {\bibfnamefont {K.}~\bibnamefont
  {Pardo}}, \bibinfo {author} {\bibfnamefont {M.}~\bibnamefont {Fishbach}},
  \bibinfo {author} {\bibfnamefont {D.~E.}\ \bibnamefont {Holz}}, \ and\
  \bibinfo {author} {\bibfnamefont {D.~N.}\ \bibnamefont {Spergel}},\ }\href
  {\doibase 10.1088/1475-7516/2018/07/048} {\bibfield  {journal} {\bibinfo
  {journal} {JCAP}\ }\textbf {\bibinfo {volume} {07}},\ \bibinfo {pages} {048}
  (\bibinfo {year} {2018})},\ \Eprint {http://arxiv.org/abs/1801.08160}
  {arXiv:1801.08160 [gr-qc]} \BibitemShut {NoStop}%
\bibitem [{\citenamefont {Abbott}\ \emph {et~al.}(2019)\citenamefont {Abbott}
  \emph {et~al.}}]{Abbott:2018lct}%
  \BibitemOpen
  \bibfield  {author} {\bibinfo {author} {\bibfnamefont {B.~P.}\ \bibnamefont
  {Abbott}} \emph {et~al.} (\bibinfo {collaboration} {LIGO Scientific,
  Virgo}),\ }\href {\doibase 10.1103/PhysRevLett.123.011102} {\bibfield
  {journal} {\bibinfo  {journal} {Phys. Rev. Lett.}\ }\textbf {\bibinfo
  {volume} {123}},\ \bibinfo {pages} {011102} (\bibinfo {year} {2019})},\
  \Eprint {http://arxiv.org/abs/1811.00364} {arXiv:1811.00364 [gr-qc]}
  \BibitemShut {NoStop}%
\bibitem [{\citenamefont {Dias}\ \emph {et~al.}(2009)\citenamefont {Dias},
  \citenamefont {Figueras}, \citenamefont {Monteiro}, \citenamefont {Santos},\
  and\ \citenamefont {Emparan}}]{Dias:2009iu}%
  \BibitemOpen
  \bibfield  {author} {\bibinfo {author} {\bibfnamefont {O.~J.~C.}\
  \bibnamefont {Dias}}, \bibinfo {author} {\bibfnamefont {P.}~\bibnamefont
  {Figueras}}, \bibinfo {author} {\bibfnamefont {R.}~\bibnamefont {Monteiro}},
  \bibinfo {author} {\bibfnamefont {J.~E.}\ \bibnamefont {Santos}}, \ and\
  \bibinfo {author} {\bibfnamefont {R.}~\bibnamefont {Emparan}},\ }\href
  {\doibase 10.1103/PhysRevD.80.111701} {\bibfield  {journal} {\bibinfo
  {journal} {Phys. Rev. D}\ }\textbf {\bibinfo {volume} {80}},\ \bibinfo
  {pages} {111701} (\bibinfo {year} {2009})},\ \Eprint
  {http://arxiv.org/abs/0907.2248} {arXiv:0907.2248 [hep-th]} \BibitemShut
  {NoStop}%
\bibitem [{\citenamefont {Emparan}\ \emph {et~al.}(2010)\citenamefont
  {Emparan}, \citenamefont {Harmark}, \citenamefont {Niarchos},\ and\
  \citenamefont {Obers}}]{Emparan:2009at}%
  \BibitemOpen
  \bibfield  {author} {\bibinfo {author} {\bibfnamefont {R.}~\bibnamefont
  {Emparan}}, \bibinfo {author} {\bibfnamefont {T.}~\bibnamefont {Harmark}},
  \bibinfo {author} {\bibfnamefont {V.}~\bibnamefont {Niarchos}}, \ and\
  \bibinfo {author} {\bibfnamefont {N.~A.}\ \bibnamefont {Obers}},\ }\href
  {\doibase 10.1007/JHEP03(2010)063} {\bibfield  {journal} {\bibinfo  {journal}
  {JHEP}\ }\textbf {\bibinfo {volume} {03}},\ \bibinfo {pages} {063} (\bibinfo
  {year} {2010})},\ \Eprint {http://arxiv.org/abs/0910.1601} {arXiv:0910.1601
  [hep-th]} \BibitemShut {NoStop}%
\bibitem [{\citenamefont {McWilliams}(2010)}]{McWilliams:2009ym}%
  \BibitemOpen
  \bibfield  {author} {\bibinfo {author} {\bibfnamefont {S.~T.}\ \bibnamefont
  {McWilliams}},\ }\href {\doibase 10.1103/PhysRevLett.104.141601} {\bibfield
  {journal} {\bibinfo  {journal} {Phys. Rev. Lett.}\ }\textbf {\bibinfo
  {volume} {104}},\ \bibinfo {pages} {141601} (\bibinfo {year} {2010})},\
  \Eprint {http://arxiv.org/abs/0912.4744} {arXiv:0912.4744 [gr-qc]}
  \BibitemShut {NoStop}%
\bibitem [{\citenamefont {Figueras}\ and\ \citenamefont
  {Wiseman}(2011)}]{Figueras:2011gd}%
  \BibitemOpen
  \bibfield  {author} {\bibinfo {author} {\bibfnamefont {P.}~\bibnamefont
  {Figueras}}\ and\ \bibinfo {author} {\bibfnamefont {T.}~\bibnamefont
  {Wiseman}},\ }\href {\doibase 10.1103/PhysRevLett.107.081101} {\bibfield
  {journal} {\bibinfo  {journal} {Phys. Rev. Lett.}\ }\textbf {\bibinfo
  {volume} {107}},\ \bibinfo {pages} {081101} (\bibinfo {year} {2011})},\
  \Eprint {http://arxiv.org/abs/1105.2558} {arXiv:1105.2558 [hep-th]}
  \BibitemShut {NoStop}%
\bibitem [{\citenamefont {Abdolrahimi}\ \emph {et~al.}(2013)\citenamefont
  {Abdolrahimi}, \citenamefont {Cattoen}, \citenamefont {Page},\ and\
  \citenamefont {Yaghoobpour-Tari}}]{Abdolrahimi:2012qi}%
  \BibitemOpen
  \bibfield  {author} {\bibinfo {author} {\bibfnamefont {S.}~\bibnamefont
  {Abdolrahimi}}, \bibinfo {author} {\bibfnamefont {C.}~\bibnamefont
  {Cattoen}}, \bibinfo {author} {\bibfnamefont {D.~N.}\ \bibnamefont {Page}}, \
  and\ \bibinfo {author} {\bibfnamefont {S.}~\bibnamefont {Yaghoobpour-Tari}},\
  }\href {\doibase 10.1016/j.physletb.2013.02.034} {\bibfield  {journal}
  {\bibinfo  {journal} {Phys. Lett. B}\ }\textbf {\bibinfo {volume} {720}},\
  \bibinfo {pages} {405} (\bibinfo {year} {2013})},\ \Eprint
  {http://arxiv.org/abs/1206.0708} {arXiv:1206.0708 [hep-th]} \BibitemShut
  {NoStop}%
\bibitem [{\citenamefont {Dine}\ and\ \citenamefont
  {Fischler}(1983)}]{Dine:1982ah}%
  \BibitemOpen
  \bibfield  {author} {\bibinfo {author} {\bibfnamefont {M.}~\bibnamefont
  {Dine}}\ and\ \bibinfo {author} {\bibfnamefont {W.}~\bibnamefont
  {Fischler}},\ }\href {\doibase 10.1016/0370-2693(83)90639-1} {\bibfield
  {journal} {\bibinfo  {journal} {Phys. Lett. B}\ }\textbf {\bibinfo {volume}
  {120}},\ \bibinfo {pages} {137} (\bibinfo {year} {1983})}\BibitemShut
  {NoStop}%
\bibitem [{\citenamefont {Dvali}(2010)}]{Dvali:2007hz}%
  \BibitemOpen
  \bibfield  {author} {\bibinfo {author} {\bibfnamefont {G.}~\bibnamefont
  {Dvali}},\ }\href {\doibase 10.1002/prop.201000009} {\bibfield  {journal}
  {\bibinfo  {journal} {Fortsch. Phys.}\ }\textbf {\bibinfo {volume} {58}},\
  \bibinfo {pages} {528} (\bibinfo {year} {2010})},\ \Eprint
  {http://arxiv.org/abs/0706.2050} {arXiv:0706.2050 [hep-th]} \BibitemShut
  {NoStop}%
\bibitem [{\citenamefont {Arvanitaki}\ \emph {et~al.}(2010)\citenamefont
  {Arvanitaki}, \citenamefont {Dimopoulos}, \citenamefont {Dubovsky},
  \citenamefont {Kaloper},\ and\ \citenamefont
  {March-Russell}}]{Arvanitaki:2009fg}%
  \BibitemOpen
  \bibfield  {author} {\bibinfo {author} {\bibfnamefont {A.}~\bibnamefont
  {Arvanitaki}}, \bibinfo {author} {\bibfnamefont {S.}~\bibnamefont
  {Dimopoulos}}, \bibinfo {author} {\bibfnamefont {S.}~\bibnamefont
  {Dubovsky}}, \bibinfo {author} {\bibfnamefont {N.}~\bibnamefont {Kaloper}}, \
  and\ \bibinfo {author} {\bibfnamefont {J.}~\bibnamefont {March-Russell}},\
  }\href {\doibase 10.1103/PhysRevD.81.123530} {\bibfield  {journal} {\bibinfo
  {journal} {Phys. Rev. D}\ }\textbf {\bibinfo {volume} {81}},\ \bibinfo
  {pages} {123530} (\bibinfo {year} {2010})},\ \Eprint
  {http://arxiv.org/abs/0905.4720} {arXiv:0905.4720 [hep-th]} \BibitemShut
  {NoStop}%
\bibitem [{\citenamefont {Davoudiasl}\ \emph {et~al.}(2021)\citenamefont
  {Davoudiasl}, \citenamefont {Denton},\ and\ \citenamefont
  {McGady}}]{Davoudiasl:2020uig}%
  \BibitemOpen
  \bibfield  {author} {\bibinfo {author} {\bibfnamefont {H.}~\bibnamefont
  {Davoudiasl}}, \bibinfo {author} {\bibfnamefont {P.~B.}\ \bibnamefont
  {Denton}}, \ and\ \bibinfo {author} {\bibfnamefont {D.~A.}\ \bibnamefont
  {McGady}},\ }\href {\doibase 10.1103/PhysRevD.103.055014} {\bibfield
  {journal} {\bibinfo  {journal} {Phys. Rev. D}\ }\textbf {\bibinfo {volume}
  {103}},\ \bibinfo {pages} {055014} (\bibinfo {year} {2021})},\ \Eprint
  {http://arxiv.org/abs/2008.06505} {arXiv:2008.06505 [hep-ph]} \BibitemShut
  {NoStop}%
\bibitem [{\citenamefont {Giddings}(2017{\natexlab{a}})}]{Giddings:2017mym}%
  \BibitemOpen
  \bibfield  {author} {\bibinfo {author} {\bibfnamefont {S.~B.}\ \bibnamefont
  {Giddings}},\ }\href {\doibase 10.1007/JHEP12(2017)047} {\bibfield  {journal}
  {\bibinfo  {journal} {JHEP}\ }\textbf {\bibinfo {volume} {12}},\ \bibinfo
  {pages} {047} (\bibinfo {year} {2017}{\natexlab{a}})},\ \Eprint
  {http://arxiv.org/abs/1701.08765} {arXiv:1701.08765 [hep-th]} \BibitemShut
  {NoStop}%
\bibitem [{\citenamefont {Giddings}(2017{\natexlab{b}})}]{Giddings:2017jts}%
  \BibitemOpen
  \bibfield  {author} {\bibinfo {author} {\bibfnamefont {S.~B.}\ \bibnamefont
  {Giddings}},\ }\href {\doibase 10.1038/s41550-017-0067} {\bibfield  {journal}
  {\bibinfo  {journal} {Nature Astron.}\ }\textbf {\bibinfo {volume} {1}},\
  \bibinfo {pages} {0067} (\bibinfo {year} {2017}{\natexlab{b}})},\ \Eprint
  {http://arxiv.org/abs/1703.03387} {arXiv:1703.03387 [gr-qc]} \BibitemShut
  {NoStop}%
\bibitem [{\citenamefont {Adler}(2004)}]{Adler:2002fu}%
  \BibitemOpen
  \bibfield  {author} {\bibinfo {author} {\bibfnamefont {S.~L.}\ \bibnamefont
  {Adler}},\ }\href@noop {} {\emph {\bibinfo {title} {{Quantum theory as an
  emergent phenomenon: The statistical mechanics of matrix models as the
  precursor of quantum field theory}}}}\ (\bibinfo  {publisher} {Cambridge
  Univ. Pr.},\ \bibinfo {address} {Cambridge, UK},\ \bibinfo {year} {2004})\
  \Eprint {http://arxiv.org/abs/hep-th/0206120} {arXiv:hep-th/0206120}
  \BibitemShut {NoStop}%
\bibitem [{\citenamefont {Adler}\ and\ \citenamefont
  {Ramazano\u{g}lu}(2017)}]{Adler:2013mna}%
  \BibitemOpen
  \bibfield  {author} {\bibinfo {author} {\bibfnamefont {S.~L.}\ \bibnamefont
  {Adler}}\ and\ \bibinfo {author} {\bibfnamefont {F.~M.}\ \bibnamefont
  {Ramazano\u{g}lu}},\ }\href {\doibase 10.1142/S021827181550011X} {\bibfield
  {journal} {\bibinfo  {journal} {Int. J. Mod. Phys. D}\ }\textbf {\bibinfo
  {volume} {24}},\ \bibinfo {pages} {1550011} (\bibinfo {year} {2017})},\
  \bibinfo {note} {[Erratum: Int.J.Mod.Phys.D 26, 1792002 (2017)]},\ \Eprint
  {http://arxiv.org/abs/1308.1448} {arXiv:1308.1448 [gr-qc]} \BibitemShut
  {NoStop}%
\bibitem [{\citenamefont {Adler}(2021)}]{Adler:2021urw}%
  \BibitemOpen
  \bibfield  {author} {\bibinfo {author} {\bibfnamefont {S.~L.}\ \bibnamefont
  {Adler}},\ }\href@noop {} {\  (\bibinfo {year} {2021})},\ \Eprint
  {http://arxiv.org/abs/2107.11816} {arXiv:2107.11816 [gr-qc]} \BibitemShut
  {NoStop}%
\bibitem [{\citenamefont {{Hadjidemetriou}}(1963)}]{1963Icar....2..440H}%
  \BibitemOpen
  \bibfield  {author} {\bibinfo {author} {\bibfnamefont {J.~D.}\ \bibnamefont
  {{Hadjidemetriou}}},\ }\href {\doibase 10.1016/0019-1035(63)90072-1}
  {\bibfield  {journal} {\bibinfo  {journal} {Icarus}\ }\textbf {\bibinfo
  {volume} {2}},\ \bibinfo {pages} {440} (\bibinfo {year} {1963})}\BibitemShut
  {NoStop}%
\bibitem [{\citenamefont {{Hadjidemetriou}}(1966)}]{1966ZA.....63..116H}%
  \BibitemOpen
  \bibfield  {author} {\bibinfo {author} {\bibfnamefont {J.~D.}\ \bibnamefont
  {{Hadjidemetriou}}},\ }\href@noop {} {\bibfield  {journal} {\bibinfo
  {journal} {Zeitschrift für Astrophysik}\ }\textbf {\bibinfo {volume} {63}},\
  \bibinfo {pages} {116} (\bibinfo {year} {1966})}\BibitemShut {NoStop}%
\bibitem [{\citenamefont {Simonetti}\ \emph {et~al.}(2011)\citenamefont
  {Simonetti}, \citenamefont {Kavic}, \citenamefont {Minic}, \citenamefont
  {Surani},\ and\ \citenamefont {Vijayan}}]{Simonetti:2010mk}%
  \BibitemOpen
  \bibfield  {author} {\bibinfo {author} {\bibfnamefont {J.~H.}\ \bibnamefont
  {Simonetti}}, \bibinfo {author} {\bibfnamefont {M.}~\bibnamefont {Kavic}},
  \bibinfo {author} {\bibfnamefont {D.}~\bibnamefont {Minic}}, \bibinfo
  {author} {\bibfnamefont {U.}~\bibnamefont {Surani}}, \ and\ \bibinfo {author}
  {\bibfnamefont {V.}~\bibnamefont {Vijayan}},\ }\href {\doibase
  10.1088/2041-8205/737/2/L28} {\bibfield  {journal} {\bibinfo  {journal}
  {Astrophys. J. Lett.}\ }\textbf {\bibinfo {volume} {737}},\ \bibinfo {pages}
  {L28} (\bibinfo {year} {2011})},\ \Eprint {http://arxiv.org/abs/1010.5245}
  {arXiv:1010.5245 [astro-ph.HE]} \BibitemShut {NoStop}%
\bibitem [{\citenamefont {Chung}\ and\ \citenamefont
  {Sakellariadou}(2020)}]{Chung:2020uqj}%
  \BibitemOpen
  \bibfield  {author} {\bibinfo {author} {\bibfnamefont {K.-W.}\ \bibnamefont
  {Chung}}\ and\ \bibinfo {author} {\bibfnamefont {M.}~\bibnamefont
  {Sakellariadou}},\ }\href@noop {} {\  (\bibinfo {year} {2020})},\ \Eprint
  {http://arxiv.org/abs/2003.09778} {arXiv:2003.09778 [gr-qc]} \BibitemShut
  {NoStop}%
\bibitem [{\citenamefont {Yagi}\ \emph {et~al.}(2011)\citenamefont {Yagi},
  \citenamefont {Tanahashi},\ and\ \citenamefont {Tanaka}}]{Yagi:2011yu}%
  \BibitemOpen
  \bibfield  {author} {\bibinfo {author} {\bibfnamefont {K.}~\bibnamefont
  {Yagi}}, \bibinfo {author} {\bibfnamefont {N.}~\bibnamefont {Tanahashi}}, \
  and\ \bibinfo {author} {\bibfnamefont {T.}~\bibnamefont {Tanaka}},\ }\href
  {\doibase 10.1103/PhysRevD.83.084036} {\bibfield  {journal} {\bibinfo
  {journal} {Phys. Rev. D}\ }\textbf {\bibinfo {volume} {83}},\ \bibinfo
  {pages} {084036} (\bibinfo {year} {2011})},\ \Eprint
  {http://arxiv.org/abs/1101.4997} {arXiv:1101.4997 [gr-qc]} \BibitemShut
  {NoStop}%
\bibitem [{\citenamefont {Perkins}\ \emph {et~al.}(2021)\citenamefont
  {Perkins}, \citenamefont {Yunes},\ and\ \citenamefont
  {Berti}}]{Perkins:2020tra}%
  \BibitemOpen
  \bibfield  {author} {\bibinfo {author} {\bibfnamefont {S.~E.}\ \bibnamefont
  {Perkins}}, \bibinfo {author} {\bibfnamefont {N.}~\bibnamefont {Yunes}}, \
  and\ \bibinfo {author} {\bibfnamefont {E.}~\bibnamefont {Berti}},\ }\href
  {\doibase 10.1103/PhysRevD.103.044024} {\bibfield  {journal} {\bibinfo
  {journal} {Phys. Rev. D}\ }\textbf {\bibinfo {volume} {103}},\ \bibinfo
  {pages} {044024} (\bibinfo {year} {2021})},\ \Eprint
  {http://arxiv.org/abs/2010.09010} {arXiv:2010.09010 [gr-qc]} \BibitemShut
  {NoStop}%
\bibitem [{\citenamefont {Psaltis}(2007)}]{Psaltis:2006de}%
  \BibitemOpen
  \bibfield  {author} {\bibinfo {author} {\bibfnamefont {D.}~\bibnamefont
  {Psaltis}},\ }\href {\doibase 10.1103/PhysRevLett.98.181101} {\bibfield
  {journal} {\bibinfo  {journal} {Phys. Rev. Lett.}\ }\textbf {\bibinfo
  {volume} {98}},\ \bibinfo {pages} {181101} (\bibinfo {year} {2007})},\
  \Eprint {http://arxiv.org/abs/astro-ph/0612611} {arXiv:astro-ph/0612611}
  \BibitemShut {NoStop}%
\bibitem [{\citenamefont {Gnedin}\ \emph {et~al.}(2009)\citenamefont {Gnedin},
  \citenamefont {Maccarone}, \citenamefont {Psaltis},\ and\ \citenamefont
  {Zepf}}]{Gnedin:2009yt}%
  \BibitemOpen
  \bibfield  {author} {\bibinfo {author} {\bibfnamefont {O.~Y.}\ \bibnamefont
  {Gnedin}}, \bibinfo {author} {\bibfnamefont {T.~J.}\ \bibnamefont
  {Maccarone}}, \bibinfo {author} {\bibfnamefont {D.}~\bibnamefont {Psaltis}},
  \ and\ \bibinfo {author} {\bibfnamefont {S.~E.}\ \bibnamefont {Zepf}},\
  }\href {\doibase 10.1088/0004-637X/705/2/L168} {\bibfield  {journal}
  {\bibinfo  {journal} {Astrophys. J. Lett.}\ }\textbf {\bibinfo {volume}
  {705}},\ \bibinfo {pages} {L168} (\bibinfo {year} {2009})},\ \Eprint
  {http://arxiv.org/abs/0906.5351} {arXiv:0906.5351 [astro-ph.CO]} \BibitemShut
  {NoStop}%
\bibitem [{\citenamefont {Steele}\ \emph {et~al.}(2014)\citenamefont {Steele},
  \citenamefont {Zepf}, \citenamefont {Maccarone}, \citenamefont {Kundu},
  \citenamefont {Rhode},\ and\ \citenamefont {Salzer}}]{steele2014composition}%
  \BibitemOpen
  \bibfield  {author} {\bibinfo {author} {\bibfnamefont {M.~M.}\ \bibnamefont
  {Steele}}, \bibinfo {author} {\bibfnamefont {S.~E.}\ \bibnamefont {Zepf}},
  \bibinfo {author} {\bibfnamefont {T.~J.}\ \bibnamefont {Maccarone}}, \bibinfo
  {author} {\bibfnamefont {A.}~\bibnamefont {Kundu}}, \bibinfo {author}
  {\bibfnamefont {K.~L.}\ \bibnamefont {Rhode}}, \ and\ \bibinfo {author}
  {\bibfnamefont {J.~J.}\ \bibnamefont {Salzer}},\ }\href@noop {} {\bibfield
  {journal} {\bibinfo  {journal} {The Astrophysical Journal}\ }\textbf
  {\bibinfo {volume} {785}},\ \bibinfo {pages} {147} (\bibinfo {year}
  {2014})}\BibitemShut {NoStop}%
\bibitem [{\citenamefont {Peters}(1964)}]{Peters:1964zz}%
  \BibitemOpen
  \bibfield  {author} {\bibinfo {author} {\bibfnamefont {P.~C.}\ \bibnamefont
  {Peters}},\ }\href {\doibase 10.1103/PhysRev.136.B1224} {\bibfield  {journal}
  {\bibinfo  {journal} {Phys. Rev.}\ }\textbf {\bibinfo {volume} {136}},\
  \bibinfo {pages} {B1224} (\bibinfo {year} {1964})}\BibitemShut {NoStop}%
\bibitem [{\citenamefont {Belczynski}\ \emph {et~al.}(2001)\citenamefont
  {Belczynski}, \citenamefont {Kalogera},\ and\ \citenamefont
  {Bulik}}]{Belczynski:2001uc}%
  \BibitemOpen
  \bibfield  {author} {\bibinfo {author} {\bibfnamefont {K.}~\bibnamefont
  {Belczynski}}, \bibinfo {author} {\bibfnamefont {V.}~\bibnamefont
  {Kalogera}}, \ and\ \bibinfo {author} {\bibfnamefont {T.}~\bibnamefont
  {Bulik}},\ }\href {\doibase 10.1086/340304} {\bibfield  {journal} {\bibinfo
  {journal} {Astrophys. J.}\ }\textbf {\bibinfo {volume} {572}},\ \bibinfo
  {pages} {407} (\bibinfo {year} {2001})},\ \Eprint
  {http://arxiv.org/abs/astro-ph/0111452} {arXiv:astro-ph/0111452} \BibitemShut
  {NoStop}%
\bibitem [{\citenamefont {Barack}\ \emph {et~al.}(2019)\citenamefont {Barack}
  \emph {et~al.}}]{Barack:2018yly}%
  \BibitemOpen
  \bibfield  {author} {\bibinfo {author} {\bibfnamefont {L.}~\bibnamefont
  {Barack}} \emph {et~al.},\ }\href {\doibase 10.1088/1361-6382/ab0587}
  {\bibfield  {journal} {\bibinfo  {journal} {Class. Quant. Grav.}\ }\textbf
  {\bibinfo {volume} {36}},\ \bibinfo {pages} {143001} (\bibinfo {year}
  {2019})},\ \Eprint {http://arxiv.org/abs/1806.05195} {arXiv:1806.05195
  [gr-qc]} \BibitemShut {NoStop}%
\bibitem [{\citenamefont {Cardoso}\ \emph {et~al.}(2021)\citenamefont
  {Cardoso}, \citenamefont {Macedo},\ and\ \citenamefont
  {Vicente}}]{Cardoso:2020iji}%
  \BibitemOpen
  \bibfield  {author} {\bibinfo {author} {\bibfnamefont {V.}~\bibnamefont
  {Cardoso}}, \bibinfo {author} {\bibfnamefont {C.~F.~B.}\ \bibnamefont
  {Macedo}}, \ and\ \bibinfo {author} {\bibfnamefont {R.}~\bibnamefont
  {Vicente}},\ }\href {\doibase 10.1103/PhysRevD.103.023015} {\bibfield
  {journal} {\bibinfo  {journal} {Phys. Rev. D}\ }\textbf {\bibinfo {volume}
  {103}},\ \bibinfo {pages} {023015} (\bibinfo {year} {2021})},\ \Eprint
  {http://arxiv.org/abs/2010.15151} {arXiv:2010.15151 [gr-qc]} \BibitemShut
  {NoStop}%
\bibitem [{\citenamefont {Agol}\ and\ \citenamefont
  {Kamionkowski}(2002)}]{Agol:2001hb}%
  \BibitemOpen
  \bibfield  {author} {\bibinfo {author} {\bibfnamefont {E.}~\bibnamefont
  {Agol}}\ and\ \bibinfo {author} {\bibfnamefont {M.}~\bibnamefont
  {Kamionkowski}},\ }\href {\doibase 10.1046/j.1365-8711.2002.05523.x}
  {\bibfield  {journal} {\bibinfo  {journal} {Mon. Not. Roy. Astron. Soc.}\
  }\textbf {\bibinfo {volume} {334}},\ \bibinfo {pages} {553} (\bibinfo {year}
  {2002})},\ \Eprint {http://arxiv.org/abs/astro-ph/0109539}
  {arXiv:astro-ph/0109539} \BibitemShut {NoStop}%
\bibitem [{\citenamefont {Dong}\ \emph {et~al.}(2016)\citenamefont {Dong},
  \citenamefont {Kinney},\ and\ \citenamefont {Stojkovic}}]{Dong:2015yjs}%
  \BibitemOpen
  \bibfield  {author} {\bibinfo {author} {\bibfnamefont {R.}~\bibnamefont
  {Dong}}, \bibinfo {author} {\bibfnamefont {W.~H.}\ \bibnamefont {Kinney}}, \
  and\ \bibinfo {author} {\bibfnamefont {D.}~\bibnamefont {Stojkovic}},\ }\href
  {\doibase 10.1088/1475-7516/2016/10/034} {\bibfield  {journal} {\bibinfo
  {journal} {JCAP}\ }\textbf {\bibinfo {volume} {10}},\ \bibinfo {pages} {034}
  (\bibinfo {year} {2016})},\ \Eprint {http://arxiv.org/abs/1511.05642}
  {arXiv:1511.05642 [astro-ph.CO]} \BibitemShut {NoStop}%
\bibitem [{\citenamefont {Aasi}\ \emph {et~al.}(2015)\citenamefont {Aasi} \emph
  {et~al.}}]{TheLIGOScientific:2014jea}%
  \BibitemOpen
  \bibfield  {author} {\bibinfo {author} {\bibfnamefont {J.}~\bibnamefont
  {Aasi}} \emph {et~al.} (\bibinfo {collaboration} {LIGO Scientific}),\ }\href
  {\doibase 10.1088/0264-9381/32/7/074001} {\bibfield  {journal} {\bibinfo
  {journal} {Class. Quant. Grav.}\ }\textbf {\bibinfo {volume} {32}},\ \bibinfo
  {pages} {074001} (\bibinfo {year} {2015})},\ \Eprint
  {http://arxiv.org/abs/1411.4547} {arXiv:1411.4547 [gr-qc]} \BibitemShut
  {NoStop}%
\bibitem [{\citenamefont {Ackley}\ \emph {et~al.}(2020)\citenamefont {Ackley}
  \emph {et~al.}}]{Ackley:2020atn}%
  \BibitemOpen
  \bibfield  {author} {\bibinfo {author} {\bibfnamefont {K.}~\bibnamefont
  {Ackley}} \emph {et~al.},\ }\href {\doibase 10.1017/pasa.2020.39} {\bibfield
  {journal} {\bibinfo  {journal} {Publ. Astron. Soc. Austral.}\ }\textbf
  {\bibinfo {volume} {37}},\ \bibinfo {pages} {e047} (\bibinfo {year}
  {2020})},\ \Eprint {http://arxiv.org/abs/2007.03128} {arXiv:2007.03128
  [astro-ph.HE]} \BibitemShut {NoStop}%
\bibitem [{\citenamefont {Abbott}\ \emph {et~al.}(2017)\citenamefont {Abbott}
  \emph {et~al.}}]{Evans:2016mbw}%
  \BibitemOpen
  \bibfield  {author} {\bibinfo {author} {\bibfnamefont {B.~P.}\ \bibnamefont
  {Abbott}} \emph {et~al.} (\bibinfo {collaboration} {LIGO Scientific}),\
  }\href {\doibase 10.1088/1361-6382/aa51f4} {\bibfield  {journal} {\bibinfo
  {journal} {Class. Quant. Grav.}\ }\textbf {\bibinfo {volume} {34}},\ \bibinfo
  {pages} {044001} (\bibinfo {year} {2017})},\ \Eprint
  {http://arxiv.org/abs/1607.08697} {arXiv:1607.08697 [astro-ph.IM]}
  \BibitemShut {NoStop}%
\bibitem [{\citenamefont {Punturo}\ \emph {et~al.}(2010)\citenamefont {Punturo}
  \emph {et~al.}}]{Punturo:2010zz}%
  \BibitemOpen
  \bibfield  {author} {\bibinfo {author} {\bibfnamefont {M.}~\bibnamefont
  {Punturo}} \emph {et~al.},\ }\href {\doibase 10.1088/0264-9381/27/19/194002}
  {\bibfield  {journal} {\bibinfo  {journal} {Class. Quant. Grav.}\ }\textbf
  {\bibinfo {volume} {27}},\ \bibinfo {pages} {194002} (\bibinfo {year}
  {2010})}\BibitemShut {NoStop}%
\bibitem [{\citenamefont {Maggiore}\ \emph {et~al.}(2020)\citenamefont
  {Maggiore} \emph {et~al.}}]{Maggiore:2019uih}%
  \BibitemOpen
  \bibfield  {author} {\bibinfo {author} {\bibfnamefont {M.}~\bibnamefont
  {Maggiore}} \emph {et~al.},\ }\href {\doibase 10.1088/1475-7516/2020/03/050}
  {\bibfield  {journal} {\bibinfo  {journal} {JCAP}\ }\textbf {\bibinfo
  {volume} {03}},\ \bibinfo {pages} {050} (\bibinfo {year} {2020})},\ \Eprint
  {http://arxiv.org/abs/1912.02622} {arXiv:1912.02622 [astro-ph.CO]}
  \BibitemShut {NoStop}%
\bibitem [{\citenamefont {Hassan}\ \emph {et~al.}(2012)\citenamefont {Hassan},
  \citenamefont {Schmidt-May},\ and\ \citenamefont {von
  Strauss}}]{Hassan:2012wt}%
  \BibitemOpen
  \bibfield  {author} {\bibinfo {author} {\bibfnamefont {S.~F.}\ \bibnamefont
  {Hassan}}, \bibinfo {author} {\bibfnamefont {A.}~\bibnamefont {Schmidt-May}},
  \ and\ \bibinfo {author} {\bibfnamefont {M.}~\bibnamefont {von Strauss}},\
  }\href@noop {} {\  (\bibinfo {year} {2012})},\ \Eprint
  {http://arxiv.org/abs/1204.5202} {arXiv:1204.5202 [hep-th]} \BibitemShut
  {NoStop}%
\bibitem [{\citenamefont {Hinterbichler}\ and\ \citenamefont
  {Rosen}(2012)}]{Hinterbichler:2012cn}%
  \BibitemOpen
  \bibfield  {author} {\bibinfo {author} {\bibfnamefont {K.}~\bibnamefont
  {Hinterbichler}}\ and\ \bibinfo {author} {\bibfnamefont {R.~A.}\ \bibnamefont
  {Rosen}},\ }\href {\doibase 10.1007/JHEP07(2012)047} {\bibfield  {journal}
  {\bibinfo  {journal} {JHEP}\ }\textbf {\bibinfo {volume} {07}},\ \bibinfo
  {pages} {047} (\bibinfo {year} {2012})},\ \Eprint
  {http://arxiv.org/abs/1203.5783} {arXiv:1203.5783 [hep-th]} \BibitemShut
  {NoStop}%
\bibitem [{\citenamefont {Noller}\ and\ \citenamefont
  {Melville}(2015)}]{Noller:2014sta}%
  \BibitemOpen
  \bibfield  {author} {\bibinfo {author} {\bibfnamefont {J.}~\bibnamefont
  {Noller}}\ and\ \bibinfo {author} {\bibfnamefont {S.}~\bibnamefont
  {Melville}},\ }\href {\doibase 10.1088/1475-7516/2015/01/003} {\bibfield
  {journal} {\bibinfo  {journal} {JCAP}\ }\textbf {\bibinfo {volume} {01}},\
  \bibinfo {pages} {003} (\bibinfo {year} {2015})},\ \Eprint
  {http://arxiv.org/abs/1408.5131} {arXiv:1408.5131 [hep-th]} \BibitemShut
  {NoStop}%
\bibitem [{\citenamefont {Phinney}(2001)}]{Phinney:2001di}%
  \BibitemOpen
  \bibfield  {author} {\bibinfo {author} {\bibfnamefont {E.~S.}\ \bibnamefont
  {Phinney}},\ }\href@noop {} {\  (\bibinfo {year} {2001})},\ \Eprint
  {http://arxiv.org/abs/astro-ph/0108028} {arXiv:astro-ph/0108028} \BibitemShut
  {NoStop}%
\bibitem [{\citenamefont {Brito}\ \emph {et~al.}(2017)\citenamefont {Brito},
  \citenamefont {Ghosh}, \citenamefont {Barausse}, \citenamefont {Berti},
  \citenamefont {Cardoso}, \citenamefont {Dvorkin}, \citenamefont {Klein},\
  and\ \citenamefont {Pani}}]{Brito:2017wnc}%
  \BibitemOpen
  \bibfield  {author} {\bibinfo {author} {\bibfnamefont {R.}~\bibnamefont
  {Brito}}, \bibinfo {author} {\bibfnamefont {S.}~\bibnamefont {Ghosh}},
  \bibinfo {author} {\bibfnamefont {E.}~\bibnamefont {Barausse}}, \bibinfo
  {author} {\bibfnamefont {E.}~\bibnamefont {Berti}}, \bibinfo {author}
  {\bibfnamefont {V.}~\bibnamefont {Cardoso}}, \bibinfo {author} {\bibfnamefont
  {I.}~\bibnamefont {Dvorkin}}, \bibinfo {author} {\bibfnamefont
  {A.}~\bibnamefont {Klein}}, \ and\ \bibinfo {author} {\bibfnamefont
  {P.}~\bibnamefont {Pani}},\ }\href {\doibase 10.1103/PhysRevLett.119.131101}
  {\bibfield  {journal} {\bibinfo  {journal} {Phys. Rev. Lett.}\ }\textbf
  {\bibinfo {volume} {119}},\ \bibinfo {pages} {131101} (\bibinfo {year}
  {2017})},\ \Eprint {http://arxiv.org/abs/1706.05097} {arXiv:1706.05097
  [gr-qc]} \BibitemShut {NoStop}%
\bibitem [{\citenamefont {Tsukada}\ \emph {et~al.}(2019)\citenamefont
  {Tsukada}, \citenamefont {Callister}, \citenamefont {Matas},\ and\
  \citenamefont {Meyers}}]{Tsukada:2018mbp}%
  \BibitemOpen
  \bibfield  {author} {\bibinfo {author} {\bibfnamefont {L.}~\bibnamefont
  {Tsukada}}, \bibinfo {author} {\bibfnamefont {T.}~\bibnamefont {Callister}},
  \bibinfo {author} {\bibfnamefont {A.}~\bibnamefont {Matas}}, \ and\ \bibinfo
  {author} {\bibfnamefont {P.}~\bibnamefont {Meyers}},\ }\href {\doibase
  10.1103/PhysRevD.99.103015} {\bibfield  {journal} {\bibinfo  {journal} {Phys.
  Rev. D}\ }\textbf {\bibinfo {volume} {99}},\ \bibinfo {pages} {103015}
  (\bibinfo {year} {2019})},\ \Eprint {http://arxiv.org/abs/1812.09622}
  {arXiv:1812.09622 [astro-ph.HE]} \BibitemShut {NoStop}%
\bibitem [{\citenamefont {Tsukada}\ \emph {et~al.}(2021)\citenamefont
  {Tsukada}, \citenamefont {Brito}, \citenamefont {East},\ and\ \citenamefont
  {Siemonsen}}]{Tsukada:2020lgt}%
  \BibitemOpen
  \bibfield  {author} {\bibinfo {author} {\bibfnamefont {L.}~\bibnamefont
  {Tsukada}}, \bibinfo {author} {\bibfnamefont {R.}~\bibnamefont {Brito}},
  \bibinfo {author} {\bibfnamefont {W.~E.}\ \bibnamefont {East}}, \ and\
  \bibinfo {author} {\bibfnamefont {N.}~\bibnamefont {Siemonsen}},\ }\href
  {\doibase 10.1103/PhysRevD.103.083005} {\bibfield  {journal} {\bibinfo
  {journal} {Phys. Rev. D}\ }\textbf {\bibinfo {volume} {103}},\ \bibinfo
  {pages} {083005} (\bibinfo {year} {2021})},\ \Eprint
  {http://arxiv.org/abs/2011.06995} {arXiv:2011.06995 [astro-ph.HE]}
  \BibitemShut {NoStop}%
\bibitem [{\citenamefont {Yuan}\ \emph {et~al.}(2021)\citenamefont {Yuan},
  \citenamefont {Brito},\ and\ \citenamefont {Cardoso}}]{Yuan:2021ebu}%
  \BibitemOpen
  \bibfield  {author} {\bibinfo {author} {\bibfnamefont {C.}~\bibnamefont
  {Yuan}}, \bibinfo {author} {\bibfnamefont {R.}~\bibnamefont {Brito}}, \ and\
  \bibinfo {author} {\bibfnamefont {V.}~\bibnamefont {Cardoso}},\ }\href@noop
  {} {\  (\bibinfo {year} {2021})},\ \Eprint {http://arxiv.org/abs/2106.00021}
  {arXiv:2106.00021 [gr-qc]} \BibitemShut {NoStop}%
\bibitem [{\citenamefont {Schaerer}(2002)}]{Schaerer:2001jc}%
  \BibitemOpen
  \bibfield  {author} {\bibinfo {author} {\bibfnamefont {D.}~\bibnamefont
  {Schaerer}},\ }\href {\doibase 10.1051/0004-6361:20011619} {\bibfield
  {journal} {\bibinfo  {journal} {Astron. Astrophys.}\ }\textbf {\bibinfo
  {volume} {382}},\ \bibinfo {pages} {28} (\bibinfo {year} {2002})},\ \Eprint
  {http://arxiv.org/abs/astro-ph/0110697} {arXiv:astro-ph/0110697} \BibitemShut
  {NoStop}%
\bibitem [{\citenamefont {Fryer}\ \emph {et~al.}(2012)\citenamefont {Fryer},
  \citenamefont {Belczynski}, \citenamefont {Wiktorowicz}, \citenamefont
  {Dominik}, \citenamefont {Kalogera},\ and\ \citenamefont
  {Holz}}]{Fryer:2011cx}%
  \BibitemOpen
  \bibfield  {author} {\bibinfo {author} {\bibfnamefont {C.~L.}\ \bibnamefont
  {Fryer}}, \bibinfo {author} {\bibfnamefont {K.}~\bibnamefont {Belczynski}},
  \bibinfo {author} {\bibfnamefont {G.}~\bibnamefont {Wiktorowicz}}, \bibinfo
  {author} {\bibfnamefont {M.}~\bibnamefont {Dominik}}, \bibinfo {author}
  {\bibfnamefont {V.}~\bibnamefont {Kalogera}}, \ and\ \bibinfo {author}
  {\bibfnamefont {D.~E.}\ \bibnamefont {Holz}},\ }\href {\doibase
  10.1088/0004-637X/749/1/91} {\bibfield  {journal} {\bibinfo  {journal}
  {Astrophys. J.}\ }\textbf {\bibinfo {volume} {749}},\ \bibinfo {pages} {91}
  (\bibinfo {year} {2012})},\ \Eprint {http://arxiv.org/abs/1110.1726}
  {arXiv:1110.1726 [astro-ph.SR]} \BibitemShut {NoStop}%
\bibitem [{\citenamefont {Belczynski}\ \emph {et~al.}(2016)\citenamefont
  {Belczynski}, \citenamefont {Holz}, \citenamefont {Bulik},\ and\
  \citenamefont {O'Shaughnessy}}]{Belczynski:2016obo}%
  \BibitemOpen
  \bibfield  {author} {\bibinfo {author} {\bibfnamefont {K.}~\bibnamefont
  {Belczynski}}, \bibinfo {author} {\bibfnamefont {D.~E.}\ \bibnamefont
  {Holz}}, \bibinfo {author} {\bibfnamefont {T.}~\bibnamefont {Bulik}}, \ and\
  \bibinfo {author} {\bibfnamefont {R.}~\bibnamefont {O'Shaughnessy}},\ }\href
  {\doibase 10.1038/nature18322} {\bibfield  {journal} {\bibinfo  {journal}
  {Nature}\ }\textbf {\bibinfo {volume} {534}},\ \bibinfo {pages} {512}
  (\bibinfo {year} {2016})},\ \Eprint {http://arxiv.org/abs/1602.04531}
  {arXiv:1602.04531 [astro-ph.HE]} \BibitemShut {NoStop}%
\bibitem [{\citenamefont {Vagnozzi}(2019)}]{Vagnozzi:2017wge}%
  \BibitemOpen
  \bibfield  {author} {\bibinfo {author} {\bibfnamefont {S.}~\bibnamefont
  {Vagnozzi}},\ }\href {\doibase 10.3390/atoms7020041} {\bibfield  {journal}
  {\bibinfo  {journal} {Atoms}\ }\textbf {\bibinfo {volume} {7}},\ \bibinfo
  {pages} {41} (\bibinfo {year} {2019})},\ \Eprint
  {http://arxiv.org/abs/1703.10834} {arXiv:1703.10834 [astro-ph.SR]}
  \BibitemShut {NoStop}%
\bibitem [{\citenamefont {Springel}\ and\ \citenamefont
  {Hernquist}(2003)}]{Springel:2002ux}%
  \BibitemOpen
  \bibfield  {author} {\bibinfo {author} {\bibfnamefont {V.}~\bibnamefont
  {Springel}}\ and\ \bibinfo {author} {\bibfnamefont {L.}~\bibnamefont
  {Hernquist}},\ }\href {\doibase 10.1046/j.1365-8711.2003.06207.x} {\bibfield
  {journal} {\bibinfo  {journal} {Mon. Not. Roy. Astron. Soc.}\ }\textbf
  {\bibinfo {volume} {339}},\ \bibinfo {pages} {312} (\bibinfo {year}
  {2003})},\ \Eprint {http://arxiv.org/abs/astro-ph/0206395}
  {arXiv:astro-ph/0206395} \BibitemShut {NoStop}%
\bibitem [{\citenamefont {Vangioni}\ \emph {et~al.}(2015)\citenamefont
  {Vangioni}, \citenamefont {Olive}, \citenamefont {Prestegard}, \citenamefont
  {Silk}, \citenamefont {Petitjean},\ and\ \citenamefont
  {Mandic}}]{Vangioni:2014axa}%
  \BibitemOpen
  \bibfield  {author} {\bibinfo {author} {\bibfnamefont {E.}~\bibnamefont
  {Vangioni}}, \bibinfo {author} {\bibfnamefont {K.~A.}\ \bibnamefont {Olive}},
  \bibinfo {author} {\bibfnamefont {T.}~\bibnamefont {Prestegard}}, \bibinfo
  {author} {\bibfnamefont {J.}~\bibnamefont {Silk}}, \bibinfo {author}
  {\bibfnamefont {P.}~\bibnamefont {Petitjean}}, \ and\ \bibinfo {author}
  {\bibfnamefont {V.}~\bibnamefont {Mandic}},\ }\href {\doibase
  10.1093/mnras/stu2600} {\bibfield  {journal} {\bibinfo  {journal} {Mon. Not.
  Roy. Astron. Soc.}\ }\textbf {\bibinfo {volume} {447}},\ \bibinfo {pages}
  {2575} (\bibinfo {year} {2015})},\ \Eprint {http://arxiv.org/abs/1409.2462}
  {arXiv:1409.2462 [astro-ph.GA]} \BibitemShut {NoStop}%
\bibitem [{\citenamefont {Thrane}\ and\ \citenamefont
  {Romano}(2013)}]{Thrane:2013oya}%
  \BibitemOpen
  \bibfield  {author} {\bibinfo {author} {\bibfnamefont {E.}~\bibnamefont
  {Thrane}}\ and\ \bibinfo {author} {\bibfnamefont {J.~D.}\ \bibnamefont
  {Romano}},\ }\href {\doibase 10.1103/PhysRevD.88.124032} {\bibfield
  {journal} {\bibinfo  {journal} {Phys. Rev. D}\ }\textbf {\bibinfo {volume}
  {88}},\ \bibinfo {pages} {124032} (\bibinfo {year} {2013})},\ \Eprint
  {http://arxiv.org/abs/1310.5300} {arXiv:1310.5300 [astro-ph.IM]} \BibitemShut
  {NoStop}%
\bibitem [{\citenamefont {Allen}\ and\ \citenamefont
  {Romano}(1999)}]{Allen:1997ad}%
  \BibitemOpen
  \bibfield  {author} {\bibinfo {author} {\bibfnamefont {B.}~\bibnamefont
  {Allen}}\ and\ \bibinfo {author} {\bibfnamefont {J.~D.}\ \bibnamefont
  {Romano}},\ }\href {\doibase 10.1103/PhysRevD.59.102001} {\bibfield
  {journal} {\bibinfo  {journal} {Phys. Rev. D}\ }\textbf {\bibinfo {volume}
  {59}},\ \bibinfo {pages} {102001} (\bibinfo {year} {1999})},\ \Eprint
  {http://arxiv.org/abs/gr-qc/9710117} {arXiv:gr-qc/9710117} \BibitemShut
  {NoStop}%
\bibitem [{\citenamefont {Bekenstein}(1997)}]{Bekenstein:1997bt}%
  \BibitemOpen
  \bibfield  {author} {\bibinfo {author} {\bibfnamefont {J.~D.}\ \bibnamefont
  {Bekenstein}},\ }in\ \href@noop {} {\emph {\bibinfo {booktitle} {{8th Marcel
  Grossmann Meeting on Recent Developments in Theoretical and Experimental
  General Relativity, Gravitation and Relativistic Field Theories (MG 8)}}}}\
  (\bibinfo {year} {1997})\ pp.\ \bibinfo {pages} {92--111},\ \Eprint
  {http://arxiv.org/abs/gr-qc/9710076} {arXiv:gr-qc/9710076} \BibitemShut
  {NoStop}%
\bibitem [{\citenamefont {Bekenstein}\ and\ \citenamefont
  {Mukhanov}(1995)}]{Bekenstein:1995ju}%
  \BibitemOpen
  \bibfield  {author} {\bibinfo {author} {\bibfnamefont {J.~D.}\ \bibnamefont
  {Bekenstein}}\ and\ \bibinfo {author} {\bibfnamefont {V.~F.}\ \bibnamefont
  {Mukhanov}},\ }\href {\doibase 10.1016/0370-2693(95)01148-J} {\bibfield
  {journal} {\bibinfo  {journal} {Phys. Lett. B}\ }\textbf {\bibinfo {volume}
  {360}},\ \bibinfo {pages} {7} (\bibinfo {year} {1995})},\ \Eprint
  {http://arxiv.org/abs/gr-qc/9505012} {arXiv:gr-qc/9505012} \BibitemShut
  {NoStop}%
\bibitem [{\citenamefont {Agullo}\ \emph {et~al.}(2021)\citenamefont {Agullo},
  \citenamefont {Cardoso}, \citenamefont {Rio}, \citenamefont {Maggiore},\ and\
  \citenamefont {Pullin}}]{Agullo:2020hxe}%
  \BibitemOpen
  \bibfield  {author} {\bibinfo {author} {\bibfnamefont {I.}~\bibnamefont
  {Agullo}}, \bibinfo {author} {\bibfnamefont {V.}~\bibnamefont {Cardoso}},
  \bibinfo {author} {\bibfnamefont {A.~D.}\ \bibnamefont {Rio}}, \bibinfo
  {author} {\bibfnamefont {M.}~\bibnamefont {Maggiore}}, \ and\ \bibinfo
  {author} {\bibfnamefont {J.}~\bibnamefont {Pullin}},\ }\href {\doibase
  10.1103/PhysRevLett.126.041302} {\bibfield  {journal} {\bibinfo  {journal}
  {Phys. Rev. Lett.}\ }\textbf {\bibinfo {volume} {126}},\ \bibinfo {pages}
  {041302} (\bibinfo {year} {2021})},\ \Eprint
  {http://arxiv.org/abs/2007.13761} {arXiv:2007.13761 [gr-qc]} \BibitemShut
  {NoStop}%
\bibitem [{GWS()}]{GWSC}%
  \BibitemOpen
  \href {https://github.com/bingining/GWSC.jl} {\enquote {\bibinfo {title}
  {https://github.com/bingining/gwsc.jl},}\ }\BibitemShut {NoStop}%
\end{thebibliography}%
	
\end{document}